%% file: source.tex
\documentclass[10pt]{article}

\usepackage{epsfig}
\usepackage{a4wide}
\usepackage{amstex}

\begin{document}

\def\O{{\cal O}}
\def\N{{\cal N}}
\def\>t{>_{\scriptscriptstyle{\rm T}}}
\def\enu{\epsilon_\nu}
\def\pint{\int{\d^3p\over(2\pi)^3}}
\def\gint{\int[\D g]\P[g]}
\def\nbar{\overline n}
\def\barthe{{\bar\theta}}
\def\d{{\rm d}}
\def\e{{\bf e}}
\def\x{{\bf x}}
\def\0x{\x^\smalze}
\def\k{{\bf k}}
\def\X{{\bf X}}
\def\sperpx{{x_\perp}}
\def\sperpk{{k_\perp}}
\def\sbperpk{{{\bf k}_\perp}}
\def\sbperpx{{{\bf x}_\perp}}
\def\perpx{{x_{\rm S}}}
\def\perpk{{k_{\rm S}}}
\def\bperpk{{{\bf k}_{\rm S}}}
\def\bperpx{{{\bf x}_{\rm S}}}
\def\p{{\bf p}}
\def\q{{\bf q}}
\def\zr{{\bf z}}
\def\R{{\bf R}}
\def\A{{\bf A}}
\def\v{{\bf v}}
\def\cm{{\rm cm}}
\def\l{{\bf l}}
\def\sec{{\rm sec}}
\def\Ckol{C_{Kol}}
\def\flux{\bar\epsilon}
\def\zq{{\zeta_q}}
\def\b{b_{kpq}}
\def\bun{b^{\scriptscriptstyle (1)}_{kpq}}
\def\bdu{b^{\scriptscriptstyle (2)}_{kpq}}
\def\z0q{{\zeta^{\scriptscriptstyle{0}}_q}}
\def\smalS{{\scriptscriptstyle {\rm S}}}
\def\smalze{{\scriptscriptstyle (0)}}
\def\smalI{{\scriptscriptstyle {\rm I}}}
\def\smalun{{\scriptscriptstyle (1)}}
\def\smaldu{{\scriptscriptstyle (2)}}
\def\smaltr{{\scriptscriptstyle (3)}}
\def\smalL{{\scriptscriptstyle{\rm L}}}
\def\smalD{{\scriptscriptstyle{\rm D}}}
\def\smal1n{{\scriptscriptstyle (1,n)}}
\def\smaln{{\scriptscriptstyle (n)}}
\def\smalA{{\scriptscriptstyle {\rm A}}}
\def\shell{{\tt S}}
\def\ball{{\tt B}}
\def\nav{\bar N}
\def\micron{\mu{\rm m}}
\font\brm=cmr10 at 24truept
\font\bfm=cmbx10 at 15truept
\centerline{\brm The lift on a tank-treading}
\centerline{\brm ellipsoidal cell in a shear flow}
\vskip 15pt
\centerline{PACS numbers: 47.15.Gf, 47.55.kf, 87.45.Hw}
\vskip 20pt
\centerline{Piero Olla$^\dag$}
\vskip 5pt
\centerline{ISIAtA-CNR}
\centerline{Universit\'a di Lecce}
\centerline{73100 Lecce Italy}
\vskip 30pt
\centerline{\it J. Phys. II (France)}
\centerline{\it In press}
\vfill\eject
\centerline{\brm The lift on a tank-treading}
\centerline{\brm ellipsoidal cell in a bounded shear flow}
\vskip 20pt
\centerline{Piero Olla$^\dag$}
\vskip 5pt
\centerline{ISIAtA-CNR}
\centerline{Universit\'a di Lecce}
\centerline{73100 Lecce Italy}
\vskip 20pt
\centerline{\bf Abstract}
\vskip 5pt
The lift on a strongly non-spherical vesicle in a bounded shear flow, is
studied in the case the membrane moves with a velocity, which is a linear 
function of the coordinates. The magnitude of the induced drift is calculated 
as a function of the axes lengths, of the distance from the wall, and of the 
ratio of the cell to the solvent viscosity. It appears that the main mechanism 
for lift, in the presence of tank-treading motions, is the fixed orientation 
of the vesicle with respect to the flow. Tank-treading vesicles in suspensions, 
flowing through narrow gaps under small Reynolds number conditions: 
$Re=Lv/\nu < 1$ (with $L$ the gap width, v the flow velocity and $\nu$ the 
viscosity) migrate away from the walls, with a velocity that 
is $\O(Re^{-1})$ larger than predicted by inertia.
\vskip 5cm
\noindent $^\dag$ e-mail address: olla@@tiresia.isiata.le.cnr.it
\vfill\eject
An important property of suspensions, both in 
Couette and channel flow conditions, is the ability to change their
effective viscosity, by concentration of the particles, near
the centre of the flow \cite{segre62,segre62a}. One of the most striking examples is the 
behavior 
of red cells in small blood vessels, where this phenomenon takes the name
of Fahraeus-Lindqwist effect \cite{oiknine76}.

In the case of spherical particles, it is well known that the only 
mechanism, which is able to cause a transverse migration of particles, 
is the presence of inertial corrections to the Stokes equation, which 
describes the response of the solvent to the particles \cite{saffman65,vasseur76,mclaugh93}. 
This is not true anymore, however, when 
spherical symmetry is lost. In this case, a particle can behave like   
a sort of wing and move in a direction different from that of the flow.  

Very often, in particular when one deals, as in the case of red cells
\cite{oiknine83}, with vesicles, or otherwise, with droplets of an unmiscible liquid, 
the particle shape is approximately that of an ellipsoid. If the ellipsoid is 
rigid, it will not maintain a fixed orientation, but will rotate as a whole in an
unsteady fashion, carrying on a kind of flipping motion \cite{jeffery22}. One suspects 
in this case, that spherical symmetry  is in some way recovered and that Stokes dynamics 
is not able to produce by itself a transverse drift. Things change in the case of an 
object with a compliant structure, where tank-treading motions become possible and a fixed 
orientation ensues.

Recently, the behavior of a spheroidal vesicle, undergoing a tank-treading 
motion in a bounded shear flow, has been analysed using perturbation
theory around the spherical particle case \cite{olla96}. For a spherically symmetric 
geometry, a Green function for the Stokes equation is available and this 
equation can then be solved, for arbitrary boundary conditions on the 
surface of the cell. 

For strongly non-spherical objects, however, the only tractable situation 
is that of the rigid ellipsoid, studied by Jeffery \cite{jeffery22}. Keller and Skalak 
\cite{keller82} showed that Jeffery's analysis could be generalised to situations, in 
which the boundary condition on the cell surface, is that the velocity be a generic 
linear function of the coordinates. The special case they considered, with a
cell assumed to have an ellipsoidal shape, and with the points on its 
surface moving with constant angular velocity, provided a useful approximation 
for the locally area preserving motion of a red cell membrane.

Of course, this approach solves only a part of the problem, leaving out the 
determination of the shape and internal motions of a cell in an external shear.
A self-consistent description, like the one contained
e.g. in \cite{kraus96,barthes80,pozri90}, is
however not really necessary here, at least for large distances from the walls, in
which case, no large deformations are expected.

In this paper, the theory of Jeffery, and of Keller and Skalak, is used to 
calculate the lift on a vesicle, undergoing the special kind of 
tank-treading motion studied in \cite{keller82}. Analysis carried on in \cite{olla96}, 
showed in the perturbative regime, that different kinds of tank-treading motions
produce lifts, which are essentially of the same order of magnitude. (More 
precisely, the case considered in \cite{keller82} gave a lift, which was approximately 
twice as large as that of a cell, with a membrane moving in an exactly area preserving 
fashion).

As in \cite{olla96}, the analysis is confined to a range of distances from the wall,
such that a calculation of the cell-wall interaction, based on the method of 
images is possible, and that the far field expression for the velocity
disturbance from the cell can be used. Extrapolating the results to the region 
near the wall, it appears
that a strongly non-spherical tank-treading vesicle, migrates towards the 
centre of the fluid, with a speed of the same order of the difference of 
velocity, due to the external shear, that exists across its body. This 
speed is typically much larger than the one produced by inertia.
\vskip 5pt
Consider a cell, whose shape is that of an ellipsoid, with axes $a_2\le a_1
\le a_3$, immersed in a shear flow, which, in an appropiate reference 
system $\{ x_1,x_2,x_3\}$, translating with the ellipsoid and with origin at its centre,
can be written
in the form: $\bar\v=\kappa x_2\e_3$. Indicate with $\{ x'_1,x'_2,x'_3\}$
the reference system in which the equation for the ellipsoid surface can be
written in the diagonal form:
$$
{{x'_1}^2\over a_1^2}+{{x'_2}^2\over a_2^2}+{{x'_3}^2\over a_3^2}=1.
\eqno(1)
$$
In a state of tank-treading motion, the cell will keep a fixed orientation, with
$x'_1\equiv x_1$, and $x'_3$ forming an angle $\barthe$ with 
respect to $x_3$, with $0<\barthe<\pi/4$ for $\kappa>0$ (see Fig. 1). For 
vanishing internal
viscosity, or in the almost spherical case, $\barthe\to\pi/4$, corresponding
to the strain part of the shear flow dominating the dynamics and forcing the
ellipsoid and the strain axes to get aligned. In the other limit, $\barthe=0$ 
signals the crossover to the vorticity dominated, flipping motion regime.

\begin{figure}[hbtp]\centering
\centerline{
\psfig{figure=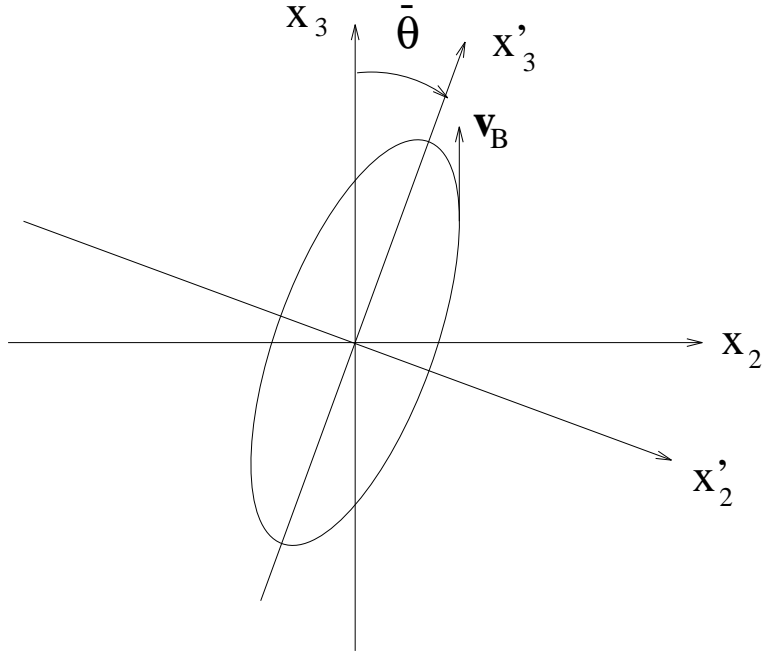,width=10.cm}
}
\caption{Orientation of a tank-treading cell in a plane shear flow $\bar\v=\kappa x_2\e_3$.}
\end{figure}

The constant angular velocity tank-treading motion, that is taken into exam,
is described by the equation for the velocity at the membrane \cite{keller82}:
$$
\v_B=\omega\kappa\Big(-{a_2\over a_3}x'_3\e_2'+{a_3\over a_2}x_2'\e_3'\Big).
\eqno(2)
$$
This kind of motion can easily be shown not to be area-preserving by an amount 
of the order of $\log a_3/a_2$; however, Eqn. (2) goes beyond providing an expression
for $\v_B$ which allows analytical calculations. Recently, Kraus et Al. \cite{kraus96} 
have obtained semi-analytical expressions, for $\barthe$ and $\omega$ in terms of 
$\v_B(x_1=0)$, 
and for $\v_B(x_1=0)$ itself, keeping into account all the couplings between the 
membrane and the fluid. It turns out that, in the spheroidal limit, the expression 
given by Eqn. (2) for $v_B(x_1=0)$, (and then also for $\barthe$ and $\omega$) is 
equal to the one obtained in \cite{kraus96}.

The boundary condition for the velocity perturbation $\v$ produced by the cell, 
is given by $\v=\bar\v-\v_B$ on the cell surface, and can be written in the form:
$$
v'_1=0,\qquad v'_2=\kappa (b x_2'+(f-\xi)x'_3),\quad 
v'_3=\kappa (-bx'_3+(f+\xi)x'_2),
\eqno(3)
$$
where:
$$
b={\sin 2\barthe\over 2},\quad{\rm and}\quad f={1\over 2}
\Big[\cos 2\barthe+\omega\Big({a_2\over a_3}-{a_3\over a_2}\Big)\Big]
\eqno(4)
$$
are the components of the strain matrix, while
$$
\xi={1\over 2}
\Big[1-\omega\Big({a_2\over a_3}+{a_3\over a_2}\Big)\Big],
\eqno(5)
$$
is the vorticity of $\v$. Jeffery's theory \cite{jeffery22} gives the following far field
expression for the velocity disturbance:
$$
\v\simeq -{4\kappa R^3\x\over x^5}\Phi(\x)
\eqno(6)
$$
where $\Phi=A{x_1'}^2+B{x'_2}^2+C{x'_3}^2+2Fx_2'x_3'$ and 
$R=(a_1a_2a_3)^{1\over 3}$. In the unprimed laboratory frame:
$$
\Phi=Ax_1^2+
\Big({B-C\over 2}\cos 2\barthe+{B+C\over 2}-F\sin 2\barthe\Big) x_2^2
$$
$$
+\Big(-{B-C\over 2}\cos 2\barthe+{B+C\over 2}+F\sin 2\barthe\Big) x_3^2
$$
$$
+[(B-C)\sin 2\barthe+2F\cos 2\barthe]x_2x_3
\eqno(7)
$$
The coefficients $A$, $B$, $C$ and $F$ are given by the expressions \cite{jeffery22}:
$$
A={(\alpha_3''-\alpha_2'')b\over Z};\qquad 
B={(2\alpha_2''+\alpha_3'')b\over Z};\qquad
C=-{(\alpha_2''+2\alpha_3'')b\over Z};
$$
$$
F={f\over 2\alpha_1'(\hat a_2^2+\hat a_3^2)}\quad{\rm where:}\quad 
Z=6(\alpha_1''\alpha_2''+\alpha_2''\alpha_3''+\alpha_3''\alpha_1''),
\eqno(8)
$$
where $\hat a_i=R^{-1}a_i$, and the 
$\alpha$-coefficients, which depend solely on the geometry of the 
cell, are given by:
$$
\alpha_1'=\int_0^\infty{\d\lambda\over 
(\hat a_2^2+\lambda)(\hat a_3^2+\lambda)\Delta},\qquad
\alpha_2'=\int_0^\infty{\d\lambda\over 
(\hat a_3^2+\lambda)(\hat a_1^2+\lambda)\Delta},...
$$
and:
$$
\alpha_1''=\int_0^\infty{\lambda\d\lambda\over 
(\hat a_2^2+\lambda)(\hat a_3^2+\lambda)\Delta},...
\eqno(9)
$$
where
$$
\Delta=\sqrt{(\hat a_1^2+\lambda)(\hat a_2^2+\lambda)(\hat a_3^2+\lambda)}.
\eqno(10)
$$
Writing $a_1=a_2=(1-\epsilon)^{1\over 2}R$, 
$a_3=(1+\epsilon)^{1\over 2}R$ and taking the limit $\epsilon\to 0$, one obtains:
$$
\Phi\simeq\sin 2\barthe [-2\hat A x_1^2+(\hat A-\tilde F)x_2^2+(\hat A+
\tilde F)x_3^2]+2(\hat F+\tilde F\cos 2\barthe)x_2x_3,
\eqno(11)
$$
where:
$$
\tilde F=-{5\epsilon\over 16}\qquad
\hat A=-{5\epsilon\over 112},\quad{\rm and}\quad
\hat F={5\over 8}\Big(1-{5\epsilon\over 14}\Big). 
\eqno(12)
$$
Upon substitution into Eqn. (6), Eqns. (11-12) give the 
expression for the velocity disturbance by a spheroidal particle, obtained 
in \cite{olla96}.

The orientation angle $\barthe$ and the tank-treading frequency $\omega$
were calculated in \cite{keller82} by Keller and Skalak, imposing angular momentum
conservation, and requiring that viscous dissipation be equal to the work done 
by $\bar\v$ on the system:  
$$
\cos 2\barthe={z_1z_3\over 4}\Big[{z_1^2\over 2}-{1\over (1-\enu)z_2-2}\Big]^{-1},
\eqno(13)
$$
$$
\omega=-{\cos 2\barthe\over z_1((1-\enu)z_2-2)}.
\eqno(14)
$$
In the equations above:
$$
z_1={1\over 2}\Big({a_3\over a_2}-{a_2\over a_3}\Big),\qquad
z_2=\alpha_1'(\hat a_3^2+\hat a_2^2),\qquad
z_3={a_3\over a_2}+{a_2\over a_3}.
\eqno(15)
$$
and $\enu$ is the ratio of the inner to outer viscosity, with the inner viscosity
being in general a weighed average of the viscosity of the membrane and that of the
fluid inside \cite{keller82}.  The situation in which the magnitude
of the RHS of Eqn. (13) is greater than one (as it happens for instance in the rigid
particle limit $\enu\to\infty$) corresponds to the flipping motion
regime. 

The derivation of Eqns. (13-15) becomes very simple when $\enu=1$
and the  membrane viscosity is equal to zero, which corresponds
to the case considered in \cite{kraus96}. It is necessary to know the 
expression of the velocity inside the cell, but it is easy to show that this is
given by the same Eqn. (2) describing 
the membrane motion (see e.g. \cite{olla96} and the expression for the inner solutions
to the Stokes equation derived there). Imposing that no energy be dissipated in 
the membrane, forces the vorticity to be continuous across the cell surface, leading
to the condition: $\xi=0$, while, requiring that no torque be present in Eqn. (3),  
leads to the strain being aligned with the ellipsoid axes: $f=0$.  These two 
conditions imply:
$v'_2=\kappa bx_2'$ and $v'_3=-\kappa bx'_3$, and using Eqns. (4-5), the result 
of Eqns. (13-15) for $\enu=1$ is recovered.

A plane wall, placed perpendicular to the $x_2$-axis, at a distance $l$ from
the cell, causes a disturbance $\v^\smalI$ in the velocity field, which, at 
the cell centre results in a net drift away from the wall. This drift can be 
calculated, if the wall is sufficiently far away, by the image method, 
imposing the no-slip boundary condition at the wall: $\v^\smalI=-\v$.
The calculation, whose details are illustrated in \cite{olla96}, is carried on 
introducing a representation in terms of potentials: $\v^\smalI=\nabla\phi
+\nabla\times\A$, $\nabla\cdot\A=0$, and Fourier transforming in $x_1$ and 
$x_3$. The potentials satisfying the Stokes equation and incompressibility
for $\v^\smalI$ are:
$$
\A_\k(x_2)=(x_2-l)\tilde\A_\k\exp(k(x_2-l)),\qquad
\phi_\k(x_2)=\tilde\phi_\k\exp(k(x_2-l)).
\eqno(16)
$$
where $f_\k(x_2)=\int\d x_1\d x_3\exp(-i\k\cdot\x)f(\x)$, $f=\{\phi,\A\}$ 
and $\k=\{ k_1,k_3\}$. 
Imposing the boundary condition on 
$\v^\smalI_\k(x_2)$, by means of Eqns. (6-7), and inverse Fourier transforming 
the result, the large distance expression for the drift $v^\smalL=-v^\smalI_2(0)$ 
is obtained:
$$
v^\smalL={U\kappa R^3\over l^2}
\eqno(17)
$$
where $U=U(\enu,r_1,r_2)$, with $r_i=a_i/a_3$, is twice the coefficient of the 
$x_2^2$ term entering $\Phi$, as given in Eqn. (7):
$$
U=(B-C)\cos 2\barthe+(B+C)-2F\sin 2\barthe.\eqno(18)
$$
In the almost spherical limit, Eqns. (11-12) lead to the result of \cite{olla96} 
$U\simeq{15\over 28}\epsilon\sin 2\barthe$, with $\barthe\simeq\pi/4$.  The 
quantity $U$, which is of order 1 for large departures from sphericity, is a 
dimensionless velocity, normalised to $\kappa R$, and gives the magnitude of the 
drift close to the wall.

Given Eqns. (6-9) and (13-15), the functional form of $U(\enu,r_1,r_2)$
can easily be explored. As it was to be expected, there is a strong
correspondence between $U(\enu,r_1,r_2)$ and $\barthe(\enu,r_1,r_2)$; in 
particular: $U(\barthe\to 0)=0$, which corresponds to the lift vanishing at 
the transition to flipping motion. 

\begin{figure}[hbtp]\centering
\centerline{
\input{fig2}
}
\caption{ Plot of the drift $U$ in function of the eccentricity
$\epsilon=({1-r_2\over 1+r_2})^{1\over 2}$, for $\enu=1$, in the case of 
axisymmetric ellipsoids: ($a$) prolate $r_1=r_2$; ($b$) oblate $r_1=1$. The 
orientation angle $\barthe$ ($c$) remains the same in the two cases.}
\end{figure}
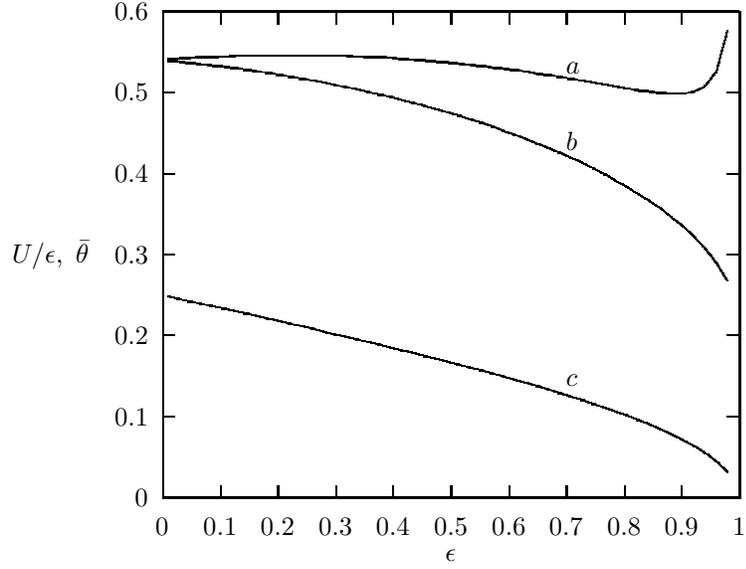

If both parameters $\enu$ and $r_2$ are kept fixed, the strongest lift is 
achieved when $r_1=r_2$ (see Fig. 2), which parallels the effect, observed in 
\cite{keller82}, that tank-treading motions are more robust in prolate rather 
than oblate ellipsoids. This is not due, however, to the orientation angle
$\barthe$, being the largest in one of the two cases. It appears instead
that $\barthe$ is independent of $r_1$: $\barthe=\barthe(\enu,r_2)$.

The strongest variations in $U$ takes place, when either the 
viscosity ratio $\enu$ or the non-sphericity (which can be parametrized by 
$r_2^{-1}$) are large. This corresponds to the intuition that strongly
non-spherical, stiff cells, prefer to stay in a flipping motion state.
In Fig. 3, as an example of strongly non spherical object, the
red cell model studied in \cite{keller82} is considered, and compared with the case of a
spheroid. The red cell, both in a stretched and an unstretched state, make a 
transition to flipping motion roughly at $\enu=3$, while the lift on a 
spheroid, remains almost constant up to $\enu\sim 10$.
For fixed $\enu$, the lift grows with the degree of non-sphericity, from 
$U=0$ at $\epsilon=0$, and reaching a maximum at a critical 
$\bar\epsilon=\bar\epsilon(\enu)$, when $r_1=r_2$. For $\epsilon>\bar\epsilon$,
$U$ decreases and reaches zero at the transition to flipping motion (see 
Fig. 4). 

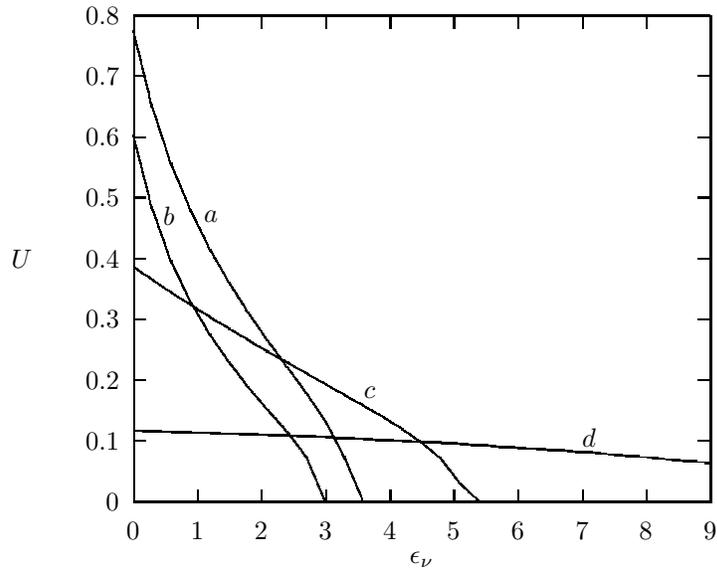
\begin{figure}[hbtp]\centering
\centerline{
\input{fig3}
}
\caption{Dependence of $U$ on the viscosity ratio
$\enu$; red cell at rest: ($a$) $r_1=1$, $r_2=0.286$; ($b$) stretched red
cell: $r_1=r_2=0.213$; ($c$) a prolate ellipsoid: $r_1=r_2=0.5$; an oblate
spheroid: $r_1=1$; ($d$) $r_2=0.8$.}
\end{figure}

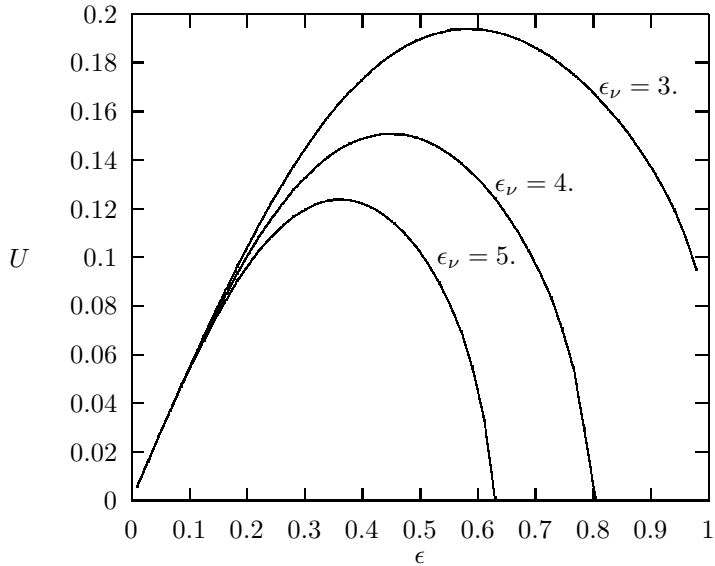
\begin{figure}[hbtp]\centering
\centerline{
\input{fig4}
 }
\caption{Dependence of $U$ on $\epsilon$, in the case of  prolate ellipsoids, for
different values of $\enu$. Notice the maxima moving towards $\epsilon=0$ as
the viscosity ratio is increased.}
\end{figure}

For $\enu$ small, it appears from Fig. 2, that the variations
in $U/\epsilon$, that take place in the case of axisymmetric ellipsoids, are
of the same order of those due to changes in the type of tank-treading 
motion. Thus, for small $\enu$, the linear theory in \cite{olla96}, 
keeps doing an acceptable job also for large values of $\epsilon$. Notice
however, that the linear theory remains unable to make predictions on 
$\barthe$, which instead, is kept fixed at $\barthe=\pi/4$.

\begin{figure}[hbtp]\centering
\centerline{
\input{fig5}
}
\caption{Values of the lift on a red cell in solvents of different viscosity,
in function of the stretching: $r_2/r_1$.}
\end{figure}
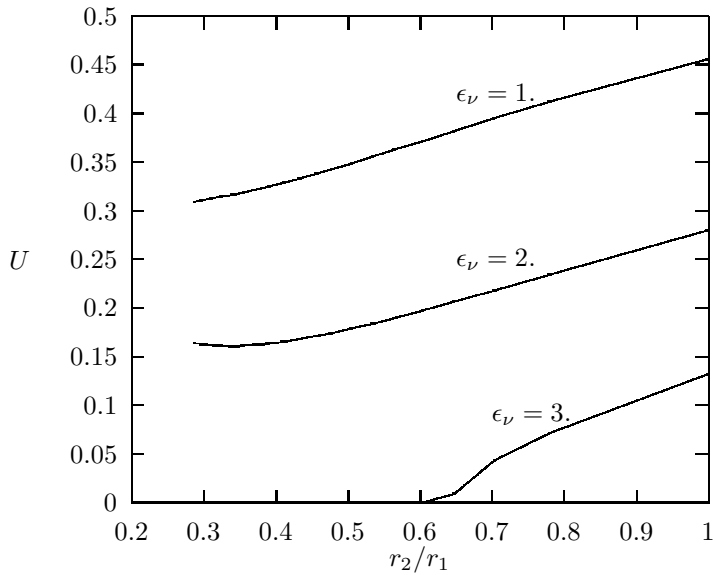

In the case of the red cell model studied in \cite{keller82}, Fig. 5 illustrates the 
dependence of the lift on the degree of stretching and on $\enu$. 
It must be remembered that, in that case, the ratio $\enu$ must account for
the viscosity of the membrane as well as that of the hemoglobin inside \cite{chien78}.
However, for $\enu\simeq 1$, corresponding to red cells in a strongly viscous
solvent, one sees that $U\simeq 0.3$, with not much 
variation when the cell is stretched by the shear. One can compare
the prediction of Eqn. (17), with the contribution to lift from inertia, 
which is: $v^\smalL\sim{f_0\kappa^2R^3\over\nu}$, with $\nu$ the viscosity of 
the suspension, and $f_0$ of the order of a few tenths \cite{vasseur76}.
Thus, if one has a suspension of red cells in a viscous solvent, forced to 
flow in a gap of thickness $L$, the condition for the lift on the red cells,
to be dominated by inertia, becomes $\kappa L^2>\nu$, i.e. a large
Reynolds number condition for the flow in the gap.

From an experimental point of view, this means that the time $T\sim L/v^\smalL$ 
for cells to concentrate near the centre of a channel flow, as revealed e.g. 
through viscometric measurement, will obey different laws in the two regimes, 
so that $T=T(\kappa)$ must have a kink at $\bar\kappa\sim\nu L^{-2}$, with 
$T\propto\kappa$ for $\kappa<\bar\kappa$ and $T\propto\kappa^2$ for 
$\kappa>\bar\kappa$. These experiments are complicated however by the 
contribution of vesicle deformability to the suspension effective viscosity 
\cite{drochon90}.

\vskip 5pt
The goal of this research was to extend the analysis carried on in \cite{olla96} to a 
non-perturbative regime. It appears that the drift produced by tank-treading 
motions approaches, close to the wall, the magnitude of the velocity difference 
at the particle scale $\kappa R$. It must be stressed, that in this region,  
the approximations leading to Eqn. (17) break up, so that the expression for the 
drift this equation provides, has a value only as an order of magnitude estimate. 
For larger values of $l/R$, however, Eqn. (17) becomes exact. This
furnishes then, a test for numerical solvers, once a boundary condition 
like that provided by Eqn. (2) is imposed.
For $l\sim R$, a more quantitative analysis would require a full solution of the 
fluid equations, in the presence of both the cell and the wall. Due to the complicated 
geometry of the problem, a numerical approach would become necessary. 

Looking at the problem from a different angle, analysis carried on in \cite{olla96} 
suggested a rather strong dependence of the drift on the exact nature of the 
tank-treading motion, and similarly, on deviations from a strictly ellipsoidal 
cell shape. The magnitude of the drift appears to be strongly influenced, also by 
the viscosity of the membrane and that of the fluid inside the cell. Hence,
considering both measurement difficulties, and how details in shape and membrane
motion are affected by perturbations, an exact knowledge of the value of $U$
close to the wall is perhaps not very interesting, and the use of a simple expression
for $\v_B$, like the one provided by Eqn. (2), is appropiate.

As an aside, these results imply that, a cell capable of producing tank-treading 
motions, has at its disposal a propulsion mechanism, which, beyond obtaining 
most of the necessary energy from the external flow, can be controlled by small
changes in cell shape. A similar mechanism for cell propulsion has been suggested in
\cite{stone96}.

Despite these strong dependencies, one expects the production of lift from 
the fixed orientation and non-sphericity of the cell, to be in itself a robust 
phenomenon. In particular, the same effect should be present, as assumed earlier,
even when the large
distance condition is relaxed, and more important, when quadratic corrections in 
the shear are taken into account. (In fact, in Poiseuille and channel flow 
configurations, one would expect a stabilising effect on the cell orientation
\cite{barthes82}).

\vskip 10pt
\noindent{\bf Aknowledgements}: Part of this research was carried on at CRS4,
(Cagliari) and at LFTC (University of Pernambuco, Recife). I would like 
to thank Gianluigi Zanetti and Giovani Vasconcelos for hospitality.

\vskip 20pt

\end{document}

%% file: fig2.tex
\setlength{\unitlength}{0.240900pt}
\ifx\plotpoint\undefined\newsavebox{\plotpoint}\fi
\sbox{\plotpoint}{\rule[-0.200pt]{0.400pt}{0.400pt}}%
\begin{picture}(1190,900)(0,0)
\font\gnuplot=cmr10 at 10pt
\gnuplot
\sbox{\plotpoint}{\rule[-0.200pt]{0.400pt}{0.400pt}}%
\put(220.0,113.0){\rule[-0.200pt]{218.255pt}{0.400pt}}
\put(220.0,113.0){\rule[-0.200pt]{0.400pt}{184.048pt}}
\put(220.0,113.0){\rule[-0.200pt]{4.818pt}{0.400pt}}
\put(198,113){\makebox(0,0)[r]{0}}
\put(1106.0,113.0){\rule[-0.200pt]{4.818pt}{0.400pt}}
\put(220.0,240.0){\rule[-0.200pt]{4.818pt}{0.400pt}}
\put(198,240){\makebox(0,0)[r]{0.1}}
\put(1106.0,240.0){\rule[-0.200pt]{4.818pt}{0.400pt}}
\put(220.0,368.0){\rule[-0.200pt]{4.818pt}{0.400pt}}
\put(198,368){\makebox(0,0)[r]{0.2}}
\put(1106.0,368.0){\rule[-0.200pt]{4.818pt}{0.400pt}}
\put(220.0,495.0){\rule[-0.200pt]{4.818pt}{0.400pt}}
\put(198,495){\makebox(0,0)[r]{0.3}}
\put(1106.0,495.0){\rule[-0.200pt]{4.818pt}{0.400pt}}
\put(220.0,622.0){\rule[-0.200pt]{4.818pt}{0.400pt}}
\put(198,622){\makebox(0,0)[r]{0.4}}
\put(1106.0,622.0){\rule[-0.200pt]{4.818pt}{0.400pt}}
\put(220.0,750.0){\rule[-0.200pt]{4.818pt}{0.400pt}}
\put(198,750){\makebox(0,0)[r]{0.5}}
\put(1106.0,750.0){\rule[-0.200pt]{4.818pt}{0.400pt}}
\put(220.0,877.0){\rule[-0.200pt]{4.818pt}{0.400pt}}
\put(198,877){\makebox(0,0)[r]{0.6}}
\put(1106.0,877.0){\rule[-0.200pt]{4.818pt}{0.400pt}}
\put(220.0,113.0){\rule[-0.200pt]{0.400pt}{4.818pt}}
\put(220,68){\makebox(0,0){0}}
\put(220.0,857.0){\rule[-0.200pt]{0.400pt}{4.818pt}}
\put(311.0,113.0){\rule[-0.200pt]{0.400pt}{4.818pt}}
\put(311,68){\makebox(0,0){0.1}}
\put(311.0,857.0){\rule[-0.200pt]{0.400pt}{4.818pt}}
\put(401.0,113.0){\rule[-0.200pt]{0.400pt}{4.818pt}}
\put(401,68){\makebox(0,0){0.2}}
\put(401.0,857.0){\rule[-0.200pt]{0.400pt}{4.818pt}}
\put(492.0,113.0){\rule[-0.200pt]{0.400pt}{4.818pt}}
\put(492,68){\makebox(0,0){0.3}}
\put(492.0,857.0){\rule[-0.200pt]{0.400pt}{4.818pt}}
\put(582.0,113.0){\rule[-0.200pt]{0.400pt}{4.818pt}}
\put(582,68){\makebox(0,0){0.4}}
\put(582.0,857.0){\rule[-0.200pt]{0.400pt}{4.818pt}}
\put(673.0,113.0){\rule[-0.200pt]{0.400pt}{4.818pt}}
\put(673,68){\makebox(0,0){0.5}}
\put(673.0,857.0){\rule[-0.200pt]{0.400pt}{4.818pt}}
\put(764.0,113.0){\rule[-0.200pt]{0.400pt}{4.818pt}}
\put(764,68){\makebox(0,0){0.6}}
\put(764.0,857.0){\rule[-0.200pt]{0.400pt}{4.818pt}}
\put(854.0,113.0){\rule[-0.200pt]{0.400pt}{4.818pt}}
\put(854,68){\makebox(0,0){0.7}}
\put(854.0,857.0){\rule[-0.200pt]{0.400pt}{4.818pt}}
\put(945.0,113.0){\rule[-0.200pt]{0.400pt}{4.818pt}}
\put(945,68){\makebox(0,0){0.8}}
\put(945.0,857.0){\rule[-0.200pt]{0.400pt}{4.818pt}}
\put(1035.0,113.0){\rule[-0.200pt]{0.400pt}{4.818pt}}
\put(1035,68){\makebox(0,0){0.9}}
\put(1035.0,857.0){\rule[-0.200pt]{0.400pt}{4.818pt}}
\put(1126.0,113.0){\rule[-0.200pt]{0.400pt}{4.818pt}}
\put(1126,68){\makebox(0,0){1}}
\put(1126.0,857.0){\rule[-0.200pt]{0.400pt}{4.818pt}}
\put(220.0,113.0){\rule[-0.200pt]{218.255pt}{0.400pt}}
\put(1126.0,113.0){\rule[-0.200pt]{0.400pt}{184.048pt}}
\put(220.0,877.0){\rule[-0.200pt]{218.255pt}{0.400pt}}
\put(45,495){\makebox(0,0){$U/\epsilon,\ \barthe$}}
\put(673,23){\makebox(0,0){$\epsilon$}}
\put(854,788){\makebox(0,0)[l]{$a$}}
\put(854,673){\makebox(0,0)[l]{$b$}}
\put(854,298){\makebox(0,0)[l]{$c$}}
\put(220.0,113.0){\rule[-0.200pt]{0.400pt}{184.048pt}}
\put(229,800){\usebox{\plotpoint}}
\put(229,798.17){\rule{3.700pt}{0.400pt}}
\multiput(229.00,799.17)(10.320,-2.000){2}{\rule{1.850pt}{0.400pt}}
\put(247,796.17){\rule{3.500pt}{0.400pt}}
\multiput(247.00,797.17)(9.736,-2.000){2}{\rule{1.750pt}{0.400pt}}
\put(264,794.17){\rule{3.700pt}{0.400pt}}
\multiput(264.00,795.17)(10.320,-2.000){2}{\rule{1.850pt}{0.400pt}}
\put(282,792.17){\rule{3.500pt}{0.400pt}}
\multiput(282.00,793.17)(9.736,-2.000){2}{\rule{1.750pt}{0.400pt}}
\put(299,790.17){\rule{3.700pt}{0.400pt}}
\multiput(299.00,791.17)(10.320,-2.000){2}{\rule{1.850pt}{0.400pt}}
\put(317,788.17){\rule{3.700pt}{0.400pt}}
\multiput(317.00,789.17)(10.320,-2.000){2}{\rule{1.850pt}{0.400pt}}
\multiput(335.00,786.95)(3.588,-0.447){3}{\rule{2.367pt}{0.108pt}}
\multiput(335.00,787.17)(12.088,-3.000){2}{\rule{1.183pt}{0.400pt}}
\put(352,783.17){\rule{3.700pt}{0.400pt}}
\multiput(352.00,784.17)(10.320,-2.000){2}{\rule{1.850pt}{0.400pt}}
\multiput(370.00,781.95)(3.588,-0.447){3}{\rule{2.367pt}{0.108pt}}
\multiput(370.00,782.17)(12.088,-3.000){2}{\rule{1.183pt}{0.400pt}}
\multiput(387.00,778.95)(3.811,-0.447){3}{\rule{2.500pt}{0.108pt}}
\multiput(387.00,779.17)(12.811,-3.000){2}{\rule{1.250pt}{0.400pt}}
\multiput(405.00,775.95)(3.588,-0.447){3}{\rule{2.367pt}{0.108pt}}
\multiput(405.00,776.17)(12.088,-3.000){2}{\rule{1.183pt}{0.400pt}}
\multiput(422.00,772.95)(3.811,-0.447){3}{\rule{2.500pt}{0.108pt}}
\multiput(422.00,773.17)(12.811,-3.000){2}{\rule{1.250pt}{0.400pt}}
\multiput(440.00,769.95)(3.811,-0.447){3}{\rule{2.500pt}{0.108pt}}
\multiput(440.00,770.17)(12.811,-3.000){2}{\rule{1.250pt}{0.400pt}}
\multiput(458.00,766.95)(3.588,-0.447){3}{\rule{2.367pt}{0.108pt}}
\multiput(458.00,767.17)(12.088,-3.000){2}{\rule{1.183pt}{0.400pt}}
\multiput(475.00,763.94)(2.528,-0.468){5}{\rule{1.900pt}{0.113pt}}
\multiput(475.00,764.17)(14.056,-4.000){2}{\rule{0.950pt}{0.400pt}}
\multiput(493.00,759.95)(3.588,-0.447){3}{\rule{2.367pt}{0.108pt}}
\multiput(493.00,760.17)(12.088,-3.000){2}{\rule{1.183pt}{0.400pt}}
\multiput(510.00,756.94)(2.528,-0.468){5}{\rule{1.900pt}{0.113pt}}
\multiput(510.00,757.17)(14.056,-4.000){2}{\rule{0.950pt}{0.400pt}}
\multiput(528.00,752.94)(2.382,-0.468){5}{\rule{1.800pt}{0.113pt}}
\multiput(528.00,753.17)(13.264,-4.000){2}{\rule{0.900pt}{0.400pt}}
\multiput(545.00,748.94)(2.528,-0.468){5}{\rule{1.900pt}{0.113pt}}
\multiput(545.00,749.17)(14.056,-4.000){2}{\rule{0.950pt}{0.400pt}}
\multiput(563.00,744.94)(2.528,-0.468){5}{\rule{1.900pt}{0.113pt}}
\multiput(563.00,745.17)(14.056,-4.000){2}{\rule{0.950pt}{0.400pt}}
\multiput(581.00,740.93)(1.823,-0.477){7}{\rule{1.460pt}{0.115pt}}
\multiput(581.00,741.17)(13.970,-5.000){2}{\rule{0.730pt}{0.400pt}}
\multiput(598.00,735.94)(2.528,-0.468){5}{\rule{1.900pt}{0.113pt}}
\multiput(598.00,736.17)(14.056,-4.000){2}{\rule{0.950pt}{0.400pt}}
\multiput(616.00,731.93)(1.823,-0.477){7}{\rule{1.460pt}{0.115pt}}
\multiput(616.00,732.17)(13.970,-5.000){2}{\rule{0.730pt}{0.400pt}}
\multiput(633.00,726.93)(1.935,-0.477){7}{\rule{1.540pt}{0.115pt}}
\multiput(633.00,727.17)(14.804,-5.000){2}{\rule{0.770pt}{0.400pt}}
\multiput(651.00,721.93)(1.935,-0.477){7}{\rule{1.540pt}{0.115pt}}
\multiput(651.00,722.17)(14.804,-5.000){2}{\rule{0.770pt}{0.400pt}}
\multiput(669.00,716.93)(1.823,-0.477){7}{\rule{1.460pt}{0.115pt}}
\multiput(669.00,717.17)(13.970,-5.000){2}{\rule{0.730pt}{0.400pt}}
\multiput(686.00,711.93)(1.575,-0.482){9}{\rule{1.300pt}{0.116pt}}
\multiput(686.00,712.17)(15.302,-6.000){2}{\rule{0.650pt}{0.400pt}}
\multiput(704.00,705.93)(1.823,-0.477){7}{\rule{1.460pt}{0.115pt}}
\multiput(704.00,706.17)(13.970,-5.000){2}{\rule{0.730pt}{0.400pt}}
\multiput(721.00,700.93)(1.575,-0.482){9}{\rule{1.300pt}{0.116pt}}
\multiput(721.00,701.17)(15.302,-6.000){2}{\rule{0.650pt}{0.400pt}}
\multiput(739.00,694.93)(1.255,-0.485){11}{\rule{1.071pt}{0.117pt}}
\multiput(739.00,695.17)(14.776,-7.000){2}{\rule{0.536pt}{0.400pt}}
\multiput(756.00,687.93)(1.575,-0.482){9}{\rule{1.300pt}{0.116pt}}
\multiput(756.00,688.17)(15.302,-6.000){2}{\rule{0.650pt}{0.400pt}}
\multiput(774.00,681.93)(1.332,-0.485){11}{\rule{1.129pt}{0.117pt}}
\multiput(774.00,682.17)(15.658,-7.000){2}{\rule{0.564pt}{0.400pt}}
\multiput(792.00,674.93)(1.255,-0.485){11}{\rule{1.071pt}{0.117pt}}
\multiput(792.00,675.17)(14.776,-7.000){2}{\rule{0.536pt}{0.400pt}}
\multiput(809.00,667.93)(1.332,-0.485){11}{\rule{1.129pt}{0.117pt}}
\multiput(809.00,668.17)(15.658,-7.000){2}{\rule{0.564pt}{0.400pt}}
\multiput(827.00,660.93)(1.255,-0.485){11}{\rule{1.071pt}{0.117pt}}
\multiput(827.00,661.17)(14.776,-7.000){2}{\rule{0.536pt}{0.400pt}}
\multiput(844.00,653.93)(1.154,-0.488){13}{\rule{1.000pt}{0.117pt}}
\multiput(844.00,654.17)(15.924,-8.000){2}{\rule{0.500pt}{0.400pt}}
\multiput(862.00,645.93)(1.154,-0.488){13}{\rule{1.000pt}{0.117pt}}
\multiput(862.00,646.17)(15.924,-8.000){2}{\rule{0.500pt}{0.400pt}}
\multiput(880.00,637.93)(0.961,-0.489){15}{\rule{0.856pt}{0.118pt}}
\multiput(880.00,638.17)(15.224,-9.000){2}{\rule{0.428pt}{0.400pt}}
\multiput(897.00,628.93)(1.019,-0.489){15}{\rule{0.900pt}{0.118pt}}
\multiput(897.00,629.17)(16.132,-9.000){2}{\rule{0.450pt}{0.400pt}}
\multiput(915.00,619.92)(0.860,-0.491){17}{\rule{0.780pt}{0.118pt}}
\multiput(915.00,620.17)(15.381,-10.000){2}{\rule{0.390pt}{0.400pt}}
\multiput(932.00,609.92)(0.912,-0.491){17}{\rule{0.820pt}{0.118pt}}
\multiput(932.00,610.17)(16.298,-10.000){2}{\rule{0.410pt}{0.400pt}}
\multiput(950.00,599.92)(0.779,-0.492){19}{\rule{0.718pt}{0.118pt}}
\multiput(950.00,600.17)(15.509,-11.000){2}{\rule{0.359pt}{0.400pt}}
\multiput(967.00,588.92)(0.826,-0.492){19}{\rule{0.755pt}{0.118pt}}
\multiput(967.00,589.17)(16.434,-11.000){2}{\rule{0.377pt}{0.400pt}}
\multiput(985.00,577.92)(0.755,-0.492){21}{\rule{0.700pt}{0.119pt}}
\multiput(985.00,578.17)(16.547,-12.000){2}{\rule{0.350pt}{0.400pt}}
\multiput(1003.00,565.92)(0.607,-0.494){25}{\rule{0.586pt}{0.119pt}}
\multiput(1003.00,566.17)(15.784,-14.000){2}{\rule{0.293pt}{0.400pt}}
\multiput(1020.00,551.92)(0.644,-0.494){25}{\rule{0.614pt}{0.119pt}}
\multiput(1020.00,552.17)(16.725,-14.000){2}{\rule{0.307pt}{0.400pt}}
\multiput(1038.00,537.92)(0.529,-0.494){29}{\rule{0.525pt}{0.119pt}}
\multiput(1038.00,538.17)(15.910,-16.000){2}{\rule{0.263pt}{0.400pt}}
\multiput(1055.58,520.83)(0.495,-0.526){33}{\rule{0.119pt}{0.522pt}}
\multiput(1054.17,521.92)(18.000,-17.916){2}{\rule{0.400pt}{0.261pt}}
\multiput(1073.58,501.44)(0.495,-0.648){31}{\rule{0.119pt}{0.618pt}}
\multiput(1072.17,502.72)(17.000,-20.718){2}{\rule{0.400pt}{0.309pt}}
\multiput(1090.58,478.91)(0.495,-0.810){33}{\rule{0.119pt}{0.744pt}}
\multiput(1089.17,480.45)(18.000,-27.455){2}{\rule{0.400pt}{0.372pt}}
\put(229,801){\usebox{\plotpoint}}
\put(229,801.17){\rule{3.700pt}{0.400pt}}
\multiput(229.00,800.17)(10.320,2.000){2}{\rule{1.850pt}{0.400pt}}
\put(247,802.67){\rule{4.095pt}{0.400pt}}
\multiput(247.00,802.17)(8.500,1.000){2}{\rule{2.048pt}{0.400pt}}
\put(264,803.67){\rule{4.336pt}{0.400pt}}
\multiput(264.00,803.17)(9.000,1.000){2}{\rule{2.168pt}{0.400pt}}
\put(299,804.67){\rule{4.336pt}{0.400pt}}
\multiput(299.00,804.17)(9.000,1.000){2}{\rule{2.168pt}{0.400pt}}
\put(317,805.67){\rule{4.336pt}{0.400pt}}
\multiput(317.00,805.17)(9.000,1.000){2}{\rule{2.168pt}{0.400pt}}
\put(282.0,805.0){\rule[-0.200pt]{4.095pt}{0.400pt}}
\put(352,806.67){\rule{4.336pt}{0.400pt}}
\multiput(352.00,806.17)(9.000,1.000){2}{\rule{2.168pt}{0.400pt}}
\put(335.0,807.0){\rule[-0.200pt]{4.095pt}{0.400pt}}
\put(475,806.67){\rule{4.336pt}{0.400pt}}
\multiput(475.00,807.17)(9.000,-1.000){2}{\rule{2.168pt}{0.400pt}}
\put(370.0,808.0){\rule[-0.200pt]{25.294pt}{0.400pt}}
\put(510,805.67){\rule{4.336pt}{0.400pt}}
\multiput(510.00,806.17)(9.000,-1.000){2}{\rule{2.168pt}{0.400pt}}
\put(528,804.67){\rule{4.095pt}{0.400pt}}
\multiput(528.00,805.17)(8.500,-1.000){2}{\rule{2.048pt}{0.400pt}}
\put(493.0,807.0){\rule[-0.200pt]{4.095pt}{0.400pt}}
\put(563,803.67){\rule{4.336pt}{0.400pt}}
\multiput(563.00,804.17)(9.000,-1.000){2}{\rule{2.168pt}{0.400pt}}
\put(581,802.17){\rule{3.500pt}{0.400pt}}
\multiput(581.00,803.17)(9.736,-2.000){2}{\rule{1.750pt}{0.400pt}}
\put(598,800.67){\rule{4.336pt}{0.400pt}}
\multiput(598.00,801.17)(9.000,-1.000){2}{\rule{2.168pt}{0.400pt}}
\put(616,799.67){\rule{4.095pt}{0.400pt}}
\multiput(616.00,800.17)(8.500,-1.000){2}{\rule{2.048pt}{0.400pt}}
\put(633,798.17){\rule{3.700pt}{0.400pt}}
\multiput(633.00,799.17)(10.320,-2.000){2}{\rule{1.850pt}{0.400pt}}
\put(651,796.67){\rule{4.336pt}{0.400pt}}
\multiput(651.00,797.17)(9.000,-1.000){2}{\rule{2.168pt}{0.400pt}}
\put(669,795.17){\rule{3.500pt}{0.400pt}}
\multiput(669.00,796.17)(9.736,-2.000){2}{\rule{1.750pt}{0.400pt}}
\put(686,793.17){\rule{3.700pt}{0.400pt}}
\multiput(686.00,794.17)(10.320,-2.000){2}{\rule{1.850pt}{0.400pt}}
\put(704,791.17){\rule{3.500pt}{0.400pt}}
\multiput(704.00,792.17)(9.736,-2.000){2}{\rule{1.750pt}{0.400pt}}
\put(721,789.17){\rule{3.700pt}{0.400pt}}
\multiput(721.00,790.17)(10.320,-2.000){2}{\rule{1.850pt}{0.400pt}}
\put(739,787.17){\rule{3.500pt}{0.400pt}}
\multiput(739.00,788.17)(9.736,-2.000){2}{\rule{1.750pt}{0.400pt}}
\put(756,785.17){\rule{3.700pt}{0.400pt}}
\multiput(756.00,786.17)(10.320,-2.000){2}{\rule{1.850pt}{0.400pt}}
\multiput(774.00,783.95)(3.811,-0.447){3}{\rule{2.500pt}{0.108pt}}
\multiput(774.00,784.17)(12.811,-3.000){2}{\rule{1.250pt}{0.400pt}}
\multiput(792.00,780.95)(3.588,-0.447){3}{\rule{2.367pt}{0.108pt}}
\multiput(792.00,781.17)(12.088,-3.000){2}{\rule{1.183pt}{0.400pt}}
\put(809,777.17){\rule{3.700pt}{0.400pt}}
\multiput(809.00,778.17)(10.320,-2.000){2}{\rule{1.850pt}{0.400pt}}
\multiput(827.00,775.95)(3.588,-0.447){3}{\rule{2.367pt}{0.108pt}}
\multiput(827.00,776.17)(12.088,-3.000){2}{\rule{1.183pt}{0.400pt}}
\multiput(844.00,772.95)(3.811,-0.447){3}{\rule{2.500pt}{0.108pt}}
\multiput(844.00,773.17)(12.811,-3.000){2}{\rule{1.250pt}{0.400pt}}
\multiput(862.00,769.95)(3.811,-0.447){3}{\rule{2.500pt}{0.108pt}}
\multiput(862.00,770.17)(12.811,-3.000){2}{\rule{1.250pt}{0.400pt}}
\multiput(880.00,766.95)(3.588,-0.447){3}{\rule{2.367pt}{0.108pt}}
\multiput(880.00,767.17)(12.088,-3.000){2}{\rule{1.183pt}{0.400pt}}
\multiput(897.00,763.95)(3.811,-0.447){3}{\rule{2.500pt}{0.108pt}}
\multiput(897.00,764.17)(12.811,-3.000){2}{\rule{1.250pt}{0.400pt}}
\multiput(915.00,760.95)(3.588,-0.447){3}{\rule{2.367pt}{0.108pt}}
\multiput(915.00,761.17)(12.088,-3.000){2}{\rule{1.183pt}{0.400pt}}
\multiput(932.00,757.95)(3.811,-0.447){3}{\rule{2.500pt}{0.108pt}}
\multiput(932.00,758.17)(12.811,-3.000){2}{\rule{1.250pt}{0.400pt}}
\multiput(950.00,754.95)(3.588,-0.447){3}{\rule{2.367pt}{0.108pt}}
\multiput(950.00,755.17)(12.088,-3.000){2}{\rule{1.183pt}{0.400pt}}
\put(967,751.17){\rule{3.700pt}{0.400pt}}
\multiput(967.00,752.17)(10.320,-2.000){2}{\rule{1.850pt}{0.400pt}}
\put(985,749.17){\rule{3.700pt}{0.400pt}}
\multiput(985.00,750.17)(10.320,-2.000){2}{\rule{1.850pt}{0.400pt}}
\put(1003,747.67){\rule{4.095pt}{0.400pt}}
\multiput(1003.00,748.17)(8.500,-1.000){2}{\rule{2.048pt}{0.400pt}}
\put(545.0,805.0){\rule[-0.200pt]{4.336pt}{0.400pt}}
\multiput(1038.00,748.61)(3.588,0.447){3}{\rule{2.367pt}{0.108pt}}
\multiput(1038.00,747.17)(12.088,3.000){2}{\rule{1.183pt}{0.400pt}}
\multiput(1055.00,751.59)(1.019,0.489){15}{\rule{0.900pt}{0.118pt}}
\multiput(1055.00,750.17)(16.132,9.000){2}{\rule{0.450pt}{0.400pt}}
\multiput(1073.58,760.00)(0.495,0.648){31}{\rule{0.119pt}{0.618pt}}
\multiput(1072.17,760.00)(17.000,20.718){2}{\rule{0.400pt}{0.309pt}}
\multiput(1090.58,782.00)(0.495,1.831){33}{\rule{0.119pt}{1.544pt}}
\multiput(1089.17,782.00)(18.000,61.794){2}{\rule{0.400pt}{0.772pt}}
\put(1020.0,748.0){\rule[-0.200pt]{4.336pt}{0.400pt}}
\put(229,429){\usebox{\plotpoint}}
\multiput(229.00,427.94)(2.528,-0.468){5}{\rule{1.900pt}{0.113pt}}
\multiput(229.00,428.17)(14.056,-4.000){2}{\rule{0.950pt}{0.400pt}}
\multiput(247.00,423.94)(2.382,-0.468){5}{\rule{1.800pt}{0.113pt}}
\multiput(247.00,424.17)(13.264,-4.000){2}{\rule{0.900pt}{0.400pt}}
\multiput(264.00,419.94)(2.528,-0.468){5}{\rule{1.900pt}{0.113pt}}
\multiput(264.00,420.17)(14.056,-4.000){2}{\rule{0.950pt}{0.400pt}}
\multiput(282.00,415.95)(3.588,-0.447){3}{\rule{2.367pt}{0.108pt}}
\multiput(282.00,416.17)(12.088,-3.000){2}{\rule{1.183pt}{0.400pt}}
\multiput(299.00,412.94)(2.528,-0.468){5}{\rule{1.900pt}{0.113pt}}
\multiput(299.00,413.17)(14.056,-4.000){2}{\rule{0.950pt}{0.400pt}}
\multiput(317.00,408.94)(2.528,-0.468){5}{\rule{1.900pt}{0.113pt}}
\multiput(317.00,409.17)(14.056,-4.000){2}{\rule{0.950pt}{0.400pt}}
\multiput(335.00,404.94)(2.382,-0.468){5}{\rule{1.800pt}{0.113pt}}
\multiput(335.00,405.17)(13.264,-4.000){2}{\rule{0.900pt}{0.400pt}}
\multiput(352.00,400.94)(2.528,-0.468){5}{\rule{1.900pt}{0.113pt}}
\multiput(352.00,401.17)(14.056,-4.000){2}{\rule{0.950pt}{0.400pt}}
\multiput(370.00,396.94)(2.382,-0.468){5}{\rule{1.800pt}{0.113pt}}
\multiput(370.00,397.17)(13.264,-4.000){2}{\rule{0.900pt}{0.400pt}}
\multiput(387.00,392.94)(2.528,-0.468){5}{\rule{1.900pt}{0.113pt}}
\multiput(387.00,393.17)(14.056,-4.000){2}{\rule{0.950pt}{0.400pt}}
\multiput(405.00,388.94)(2.382,-0.468){5}{\rule{1.800pt}{0.113pt}}
\multiput(405.00,389.17)(13.264,-4.000){2}{\rule{0.900pt}{0.400pt}}
\multiput(422.00,384.94)(2.528,-0.468){5}{\rule{1.900pt}{0.113pt}}
\multiput(422.00,385.17)(14.056,-4.000){2}{\rule{0.950pt}{0.400pt}}
\multiput(440.00,380.94)(2.528,-0.468){5}{\rule{1.900pt}{0.113pt}}
\multiput(440.00,381.17)(14.056,-4.000){2}{\rule{0.950pt}{0.400pt}}
\multiput(458.00,376.93)(1.823,-0.477){7}{\rule{1.460pt}{0.115pt}}
\multiput(458.00,377.17)(13.970,-5.000){2}{\rule{0.730pt}{0.400pt}}
\multiput(475.00,371.94)(2.528,-0.468){5}{\rule{1.900pt}{0.113pt}}
\multiput(475.00,372.17)(14.056,-4.000){2}{\rule{0.950pt}{0.400pt}}
\multiput(493.00,367.94)(2.382,-0.468){5}{\rule{1.800pt}{0.113pt}}
\multiput(493.00,368.17)(13.264,-4.000){2}{\rule{0.900pt}{0.400pt}}
\multiput(510.00,363.94)(2.528,-0.468){5}{\rule{1.900pt}{0.113pt}}
\multiput(510.00,364.17)(14.056,-4.000){2}{\rule{0.950pt}{0.400pt}}
\multiput(528.00,359.94)(2.382,-0.468){5}{\rule{1.800pt}{0.113pt}}
\multiput(528.00,360.17)(13.264,-4.000){2}{\rule{0.900pt}{0.400pt}}
\multiput(545.00,355.94)(2.528,-0.468){5}{\rule{1.900pt}{0.113pt}}
\multiput(545.00,356.17)(14.056,-4.000){2}{\rule{0.950pt}{0.400pt}}
\multiput(563.00,351.93)(1.935,-0.477){7}{\rule{1.540pt}{0.115pt}}
\multiput(563.00,352.17)(14.804,-5.000){2}{\rule{0.770pt}{0.400pt}}
\multiput(581.00,346.94)(2.382,-0.468){5}{\rule{1.800pt}{0.113pt}}
\multiput(581.00,347.17)(13.264,-4.000){2}{\rule{0.900pt}{0.400pt}}
\multiput(598.00,342.94)(2.528,-0.468){5}{\rule{1.900pt}{0.113pt}}
\multiput(598.00,343.17)(14.056,-4.000){2}{\rule{0.950pt}{0.400pt}}
\multiput(616.00,338.93)(1.823,-0.477){7}{\rule{1.460pt}{0.115pt}}
\multiput(616.00,339.17)(13.970,-5.000){2}{\rule{0.730pt}{0.400pt}}
\multiput(633.00,333.94)(2.528,-0.468){5}{\rule{1.900pt}{0.113pt}}
\multiput(633.00,334.17)(14.056,-4.000){2}{\rule{0.950pt}{0.400pt}}
\multiput(651.00,329.93)(1.935,-0.477){7}{\rule{1.540pt}{0.115pt}}
\multiput(651.00,330.17)(14.804,-5.000){2}{\rule{0.770pt}{0.400pt}}
\multiput(669.00,324.94)(2.382,-0.468){5}{\rule{1.800pt}{0.113pt}}
\multiput(669.00,325.17)(13.264,-4.000){2}{\rule{0.900pt}{0.400pt}}
\multiput(686.00,320.93)(1.935,-0.477){7}{\rule{1.540pt}{0.115pt}}
\multiput(686.00,321.17)(14.804,-5.000){2}{\rule{0.770pt}{0.400pt}}
\multiput(704.00,315.94)(2.382,-0.468){5}{\rule{1.800pt}{0.113pt}}
\multiput(704.00,316.17)(13.264,-4.000){2}{\rule{0.900pt}{0.400pt}}
\multiput(721.00,311.93)(1.935,-0.477){7}{\rule{1.540pt}{0.115pt}}
\multiput(721.00,312.17)(14.804,-5.000){2}{\rule{0.770pt}{0.400pt}}
\multiput(739.00,306.93)(1.823,-0.477){7}{\rule{1.460pt}{0.115pt}}
\multiput(739.00,307.17)(13.970,-5.000){2}{\rule{0.730pt}{0.400pt}}
\multiput(756.00,301.93)(1.935,-0.477){7}{\rule{1.540pt}{0.115pt}}
\multiput(756.00,302.17)(14.804,-5.000){2}{\rule{0.770pt}{0.400pt}}
\multiput(774.00,296.93)(1.935,-0.477){7}{\rule{1.540pt}{0.115pt}}
\multiput(774.00,297.17)(14.804,-5.000){2}{\rule{0.770pt}{0.400pt}}
\multiput(792.00,291.93)(1.823,-0.477){7}{\rule{1.460pt}{0.115pt}}
\multiput(792.00,292.17)(13.970,-5.000){2}{\rule{0.730pt}{0.400pt}}
\multiput(809.00,286.93)(1.935,-0.477){7}{\rule{1.540pt}{0.115pt}}
\multiput(809.00,287.17)(14.804,-5.000){2}{\rule{0.770pt}{0.400pt}}
\multiput(827.00,281.93)(1.485,-0.482){9}{\rule{1.233pt}{0.116pt}}
\multiput(827.00,282.17)(14.440,-6.000){2}{\rule{0.617pt}{0.400pt}}
\multiput(844.00,275.93)(1.935,-0.477){7}{\rule{1.540pt}{0.115pt}}
\multiput(844.00,276.17)(14.804,-5.000){2}{\rule{0.770pt}{0.400pt}}
\multiput(862.00,270.93)(1.575,-0.482){9}{\rule{1.300pt}{0.116pt}}
\multiput(862.00,271.17)(15.302,-6.000){2}{\rule{0.650pt}{0.400pt}}
\multiput(880.00,264.93)(1.485,-0.482){9}{\rule{1.233pt}{0.116pt}}
\multiput(880.00,265.17)(14.440,-6.000){2}{\rule{0.617pt}{0.400pt}}
\multiput(897.00,258.93)(1.575,-0.482){9}{\rule{1.300pt}{0.116pt}}
\multiput(897.00,259.17)(15.302,-6.000){2}{\rule{0.650pt}{0.400pt}}
\multiput(915.00,252.93)(1.485,-0.482){9}{\rule{1.233pt}{0.116pt}}
\multiput(915.00,253.17)(14.440,-6.000){2}{\rule{0.617pt}{0.400pt}}
\multiput(932.00,246.93)(1.575,-0.482){9}{\rule{1.300pt}{0.116pt}}
\multiput(932.00,247.17)(15.302,-6.000){2}{\rule{0.650pt}{0.400pt}}
\multiput(950.00,240.93)(1.255,-0.485){11}{\rule{1.071pt}{0.117pt}}
\multiput(950.00,241.17)(14.776,-7.000){2}{\rule{0.536pt}{0.400pt}}
\multiput(967.00,233.93)(1.332,-0.485){11}{\rule{1.129pt}{0.117pt}}
\multiput(967.00,234.17)(15.658,-7.000){2}{\rule{0.564pt}{0.400pt}}
\multiput(985.00,226.93)(1.154,-0.488){13}{\rule{1.000pt}{0.117pt}}
\multiput(985.00,227.17)(15.924,-8.000){2}{\rule{0.500pt}{0.400pt}}
\multiput(1003.00,218.93)(1.088,-0.488){13}{\rule{0.950pt}{0.117pt}}
\multiput(1003.00,219.17)(15.028,-8.000){2}{\rule{0.475pt}{0.400pt}}
\multiput(1020.00,210.93)(1.019,-0.489){15}{\rule{0.900pt}{0.118pt}}
\multiput(1020.00,211.17)(16.132,-9.000){2}{\rule{0.450pt}{0.400pt}}
\multiput(1038.00,201.93)(0.961,-0.489){15}{\rule{0.856pt}{0.118pt}}
\multiput(1038.00,202.17)(15.224,-9.000){2}{\rule{0.428pt}{0.400pt}}
\multiput(1055.00,192.92)(0.826,-0.492){19}{\rule{0.755pt}{0.118pt}}
\multiput(1055.00,193.17)(16.434,-11.000){2}{\rule{0.377pt}{0.400pt}}
\multiput(1073.00,181.92)(0.655,-0.493){23}{\rule{0.623pt}{0.119pt}}
\multiput(1073.00,182.17)(15.707,-13.000){2}{\rule{0.312pt}{0.400pt}}
\multiput(1090.00,168.92)(0.528,-0.495){31}{\rule{0.524pt}{0.119pt}}
\multiput(1090.00,169.17)(16.913,-17.000){2}{\rule{0.262pt}{0.400pt}}
\end{picture}

%% file: fig3.tex
\setlength{\unitlength}{0.240900pt}
\ifx\plotpoint\undefined\newsavebox{\plotpoint}\fi
\sbox{\plotpoint}{\rule[-0.200pt]{0.400pt}{0.400pt}}%
\begin{picture}(1190,900)(0,0)
\font\gnuplot=cmr10 at 10pt
\gnuplot
\sbox{\plotpoint}{\rule[-0.200pt]{0.400pt}{0.400pt}}%
\put(220.0,113.0){\rule[-0.200pt]{218.255pt}{0.400pt}}
\put(220.0,113.0){\rule[-0.200pt]{0.400pt}{184.048pt}}
\put(220.0,113.0){\rule[-0.200pt]{4.818pt}{0.400pt}}
\put(198,113){\makebox(0,0)[r]{0}}
\put(1106.0,113.0){\rule[-0.200pt]{4.818pt}{0.400pt}}
\put(220.0,209.0){\rule[-0.200pt]{4.818pt}{0.400pt}}
\put(198,209){\makebox(0,0)[r]{0.1}}
\put(1106.0,209.0){\rule[-0.200pt]{4.818pt}{0.400pt}}
\put(220.0,304.0){\rule[-0.200pt]{4.818pt}{0.400pt}}
\put(198,304){\makebox(0,0)[r]{0.2}}
\put(1106.0,304.0){\rule[-0.200pt]{4.818pt}{0.400pt}}
\put(220.0,400.0){\rule[-0.200pt]{4.818pt}{0.400pt}}
\put(198,400){\makebox(0,0)[r]{0.3}}
\put(1106.0,400.0){\rule[-0.200pt]{4.818pt}{0.400pt}}
\put(220.0,495.0){\rule[-0.200pt]{4.818pt}{0.400pt}}
\put(198,495){\makebox(0,0)[r]{0.4}}
\put(1106.0,495.0){\rule[-0.200pt]{4.818pt}{0.400pt}}
\put(220.0,591.0){\rule[-0.200pt]{4.818pt}{0.400pt}}
\put(198,591){\makebox(0,0)[r]{0.5}}
\put(1106.0,591.0){\rule[-0.200pt]{4.818pt}{0.400pt}}
\put(220.0,686.0){\rule[-0.200pt]{4.818pt}{0.400pt}}
\put(198,686){\makebox(0,0)[r]{0.6}}
\put(1106.0,686.0){\rule[-0.200pt]{4.818pt}{0.400pt}}
\put(220.0,782.0){\rule[-0.200pt]{4.818pt}{0.400pt}}
\put(198,782){\makebox(0,0)[r]{0.7}}
\put(1106.0,782.0){\rule[-0.200pt]{4.818pt}{0.400pt}}
\put(220.0,877.0){\rule[-0.200pt]{4.818pt}{0.400pt}}
\put(198,877){\makebox(0,0)[r]{0.8}}
\put(1106.0,877.0){\rule[-0.200pt]{4.818pt}{0.400pt}}
\put(220.0,113.0){\rule[-0.200pt]{0.400pt}{4.818pt}}
\put(220,68){\makebox(0,0){0}}
\put(220.0,857.0){\rule[-0.200pt]{0.400pt}{4.818pt}}
\put(321.0,113.0){\rule[-0.200pt]{0.400pt}{4.818pt}}
\put(321,68){\makebox(0,0){1}}
\put(321.0,857.0){\rule[-0.200pt]{0.400pt}{4.818pt}}
\put(421.0,113.0){\rule[-0.200pt]{0.400pt}{4.818pt}}
\put(421,68){\makebox(0,0){2}}
\put(421.0,857.0){\rule[-0.200pt]{0.400pt}{4.818pt}}
\put(522.0,113.0){\rule[-0.200pt]{0.400pt}{4.818pt}}
\put(522,68){\makebox(0,0){3}}
\put(522.0,857.0){\rule[-0.200pt]{0.400pt}{4.818pt}}
\put(623.0,113.0){\rule[-0.200pt]{0.400pt}{4.818pt}}
\put(623,68){\makebox(0,0){4}}
\put(623.0,857.0){\rule[-0.200pt]{0.400pt}{4.818pt}}
\put(723.0,113.0){\rule[-0.200pt]{0.400pt}{4.818pt}}
\put(723,68){\makebox(0,0){5}}
\put(723.0,857.0){\rule[-0.200pt]{0.400pt}{4.818pt}}
\put(824.0,113.0){\rule[-0.200pt]{0.400pt}{4.818pt}}
\put(824,68){\makebox(0,0){6}}
\put(824.0,857.0){\rule[-0.200pt]{0.400pt}{4.818pt}}
\put(925.0,113.0){\rule[-0.200pt]{0.400pt}{4.818pt}}
\put(925,68){\makebox(0,0){7}}
\put(925.0,857.0){\rule[-0.200pt]{0.400pt}{4.818pt}}
\put(1025.0,113.0){\rule[-0.200pt]{0.400pt}{4.818pt}}
\put(1025,68){\makebox(0,0){8}}
\put(1025.0,857.0){\rule[-0.200pt]{0.400pt}{4.818pt}}
\put(1126.0,113.0){\rule[-0.200pt]{0.400pt}{4.818pt}}
\put(1126,68){\makebox(0,0){9}}
\put(1126.0,857.0){\rule[-0.200pt]{0.400pt}{4.818pt}}
\put(220.0,113.0){\rule[-0.200pt]{218.255pt}{0.400pt}}
\put(1126.0,113.0){\rule[-0.200pt]{0.400pt}{184.048pt}}
\put(220.0,877.0){\rule[-0.200pt]{218.255pt}{0.400pt}}
\put(45,495){\makebox(0,0){$U$}}
\put(673,23){\makebox(0,0){$\enu$}}
\put(331,562){\makebox(0,0)[l]{$a$}}
\put(267,562){\makebox(0,0)[l]{$b$}}
\put(582,285){\makebox(0,0)[l]{$c$}}
\put(925,214){\makebox(0,0)[l]{$d$}}
\put(220.0,113.0){\rule[-0.200pt]{0.400pt}{184.048pt}}
\put(220,690){\usebox{\plotpoint}}
\multiput(220.58,683.16)(0.497,-1.949){57}{\rule{0.120pt}{1.647pt}}
\multiput(219.17,686.58)(30.000,-112.582){2}{\rule{0.400pt}{0.823pt}}
\multiput(250.58,568.94)(0.497,-1.410){57}{\rule{0.120pt}{1.220pt}}
\multiput(249.17,571.47)(30.000,-81.468){2}{\rule{0.400pt}{0.610pt}}
\multiput(280.58,486.16)(0.497,-1.037){59}{\rule{0.120pt}{0.926pt}}
\multiput(279.17,488.08)(31.000,-62.078){2}{\rule{0.400pt}{0.463pt}}
\multiput(311.58,422.76)(0.497,-0.853){57}{\rule{0.120pt}{0.780pt}}
\multiput(310.17,424.38)(30.000,-49.381){2}{\rule{0.400pt}{0.390pt}}
\multiput(341.58,372.20)(0.497,-0.718){57}{\rule{0.120pt}{0.673pt}}
\multiput(340.17,373.60)(30.000,-41.602){2}{\rule{0.400pt}{0.337pt}}
\multiput(371.58,329.43)(0.497,-0.650){57}{\rule{0.120pt}{0.620pt}}
\multiput(370.17,330.71)(30.000,-37.713){2}{\rule{0.400pt}{0.310pt}}
\multiput(401.58,290.65)(0.497,-0.583){57}{\rule{0.120pt}{0.567pt}}
\multiput(400.17,291.82)(30.000,-33.824){2}{\rule{0.400pt}{0.283pt}}
\multiput(431.58,255.71)(0.497,-0.564){59}{\rule{0.120pt}{0.552pt}}
\multiput(430.17,256.86)(31.000,-33.855){2}{\rule{0.400pt}{0.276pt}}
\multiput(462.58,220.43)(0.497,-0.650){57}{\rule{0.120pt}{0.620pt}}
\multiput(461.17,221.71)(30.000,-37.713){2}{\rule{0.400pt}{0.310pt}}
\multiput(492.58,179.66)(0.497,-1.190){57}{\rule{0.120pt}{1.047pt}}
\multiput(491.17,181.83)(30.000,-68.828){2}{\rule{0.400pt}{0.523pt}}
\put(522.0,113.0){\rule[-0.200pt]{145.504pt}{0.400pt}}
\put(220,854){\usebox{\plotpoint}}
\multiput(220.58,846.94)(0.497,-2.017){57}{\rule{0.120pt}{1.700pt}}
\multiput(219.17,850.47)(30.000,-116.472){2}{\rule{0.400pt}{0.850pt}}
\multiput(250.58,728.55)(0.497,-1.528){57}{\rule{0.120pt}{1.313pt}}
\multiput(249.17,731.27)(30.000,-88.274){2}{\rule{0.400pt}{0.657pt}}
\multiput(280.58,638.67)(0.497,-1.184){59}{\rule{0.120pt}{1.042pt}}
\multiput(279.17,640.84)(31.000,-70.837){2}{\rule{0.400pt}{0.521pt}}
\multiput(311.58,566.21)(0.497,-1.022){57}{\rule{0.120pt}{0.913pt}}
\multiput(310.17,568.10)(30.000,-59.104){2}{\rule{0.400pt}{0.457pt}}
\multiput(341.58,505.71)(0.497,-0.870){57}{\rule{0.120pt}{0.793pt}}
\multiput(340.17,507.35)(30.000,-50.353){2}{\rule{0.400pt}{0.397pt}}
\multiput(371.58,453.98)(0.497,-0.785){57}{\rule{0.120pt}{0.727pt}}
\multiput(370.17,455.49)(30.000,-45.492){2}{\rule{0.400pt}{0.363pt}}
\multiput(401.58,407.20)(0.497,-0.718){57}{\rule{0.120pt}{0.673pt}}
\multiput(400.17,408.60)(30.000,-41.602){2}{\rule{0.400pt}{0.337pt}}
\multiput(431.58,364.34)(0.497,-0.678){59}{\rule{0.120pt}{0.642pt}}
\multiput(430.17,365.67)(31.000,-40.668){2}{\rule{0.400pt}{0.321pt}}
\multiput(462.58,322.32)(0.497,-0.684){57}{\rule{0.120pt}{0.647pt}}
\multiput(461.17,323.66)(30.000,-39.658){2}{\rule{0.400pt}{0.323pt}}
\multiput(492.58,281.15)(0.497,-0.735){57}{\rule{0.120pt}{0.687pt}}
\multiput(491.17,282.57)(30.000,-42.575){2}{\rule{0.400pt}{0.343pt}}
\multiput(522.58,236.43)(0.497,-0.954){57}{\rule{0.120pt}{0.860pt}}
\multiput(521.17,238.22)(30.000,-55.215){2}{\rule{0.400pt}{0.430pt}}
\multiput(552.58,178.71)(0.497,-1.173){57}{\rule{0.120pt}{1.033pt}}
\multiput(551.17,180.86)(30.000,-67.855){2}{\rule{0.400pt}{0.517pt}}
\put(582.0,113.0){\rule[-0.200pt]{131.050pt}{0.400pt}}
\put(220,483){\usebox{\plotpoint}}
\multiput(220.00,481.92)(0.716,-0.496){39}{\rule{0.671pt}{0.119pt}}
\multiput(220.00,482.17)(28.606,-21.000){2}{\rule{0.336pt}{0.400pt}}
\multiput(250.00,460.92)(0.753,-0.496){37}{\rule{0.700pt}{0.119pt}}
\multiput(250.00,461.17)(28.547,-20.000){2}{\rule{0.350pt}{0.400pt}}
\multiput(280.00,440.92)(0.778,-0.496){37}{\rule{0.720pt}{0.119pt}}
\multiput(280.00,441.17)(29.506,-20.000){2}{\rule{0.360pt}{0.400pt}}
\multiput(311.00,420.92)(0.793,-0.495){35}{\rule{0.732pt}{0.119pt}}
\multiput(311.00,421.17)(28.482,-19.000){2}{\rule{0.366pt}{0.400pt}}
\multiput(341.00,401.92)(0.838,-0.495){33}{\rule{0.767pt}{0.119pt}}
\multiput(341.00,402.17)(28.409,-18.000){2}{\rule{0.383pt}{0.400pt}}
\multiput(371.00,383.92)(0.838,-0.495){33}{\rule{0.767pt}{0.119pt}}
\multiput(371.00,384.17)(28.409,-18.000){2}{\rule{0.383pt}{0.400pt}}
\multiput(401.00,365.92)(0.838,-0.495){33}{\rule{0.767pt}{0.119pt}}
\multiput(401.00,366.17)(28.409,-18.000){2}{\rule{0.383pt}{0.400pt}}
\multiput(431.00,347.92)(0.919,-0.495){31}{\rule{0.829pt}{0.119pt}}
\multiput(431.00,348.17)(29.279,-17.000){2}{\rule{0.415pt}{0.400pt}}
\multiput(462.00,330.92)(0.888,-0.495){31}{\rule{0.806pt}{0.119pt}}
\multiput(462.00,331.17)(28.327,-17.000){2}{\rule{0.403pt}{0.400pt}}
\multiput(492.00,313.92)(0.888,-0.495){31}{\rule{0.806pt}{0.119pt}}
\multiput(492.00,314.17)(28.327,-17.000){2}{\rule{0.403pt}{0.400pt}}
\multiput(522.00,296.92)(0.888,-0.495){31}{\rule{0.806pt}{0.119pt}}
\multiput(522.00,297.17)(28.327,-17.000){2}{\rule{0.403pt}{0.400pt}}
\multiput(552.00,279.92)(0.888,-0.495){31}{\rule{0.806pt}{0.119pt}}
\multiput(552.00,280.17)(28.327,-17.000){2}{\rule{0.403pt}{0.400pt}}
\multiput(582.00,262.92)(0.866,-0.495){33}{\rule{0.789pt}{0.119pt}}
\multiput(582.00,263.17)(29.363,-18.000){2}{\rule{0.394pt}{0.400pt}}
\multiput(613.00,244.92)(0.793,-0.495){35}{\rule{0.732pt}{0.119pt}}
\multiput(613.00,245.17)(28.482,-19.000){2}{\rule{0.366pt}{0.400pt}}
\multiput(643.00,225.92)(0.716,-0.496){39}{\rule{0.671pt}{0.119pt}}
\multiput(643.00,226.17)(28.606,-21.000){2}{\rule{0.336pt}{0.400pt}}
\multiput(673.00,204.92)(0.600,-0.497){47}{\rule{0.580pt}{0.120pt}}
\multiput(673.00,205.17)(28.796,-25.000){2}{\rule{0.290pt}{0.400pt}}
\multiput(703.58,178.43)(0.497,-0.650){57}{\rule{0.120pt}{0.620pt}}
\multiput(702.17,179.71)(30.000,-37.713){2}{\rule{0.400pt}{0.310pt}}
\multiput(733.00,140.92)(0.534,-0.497){55}{\rule{0.528pt}{0.120pt}}
\multiput(733.00,141.17)(29.905,-29.000){2}{\rule{0.264pt}{0.400pt}}
\put(764.0,113.0){\rule[-0.200pt]{87.206pt}{0.400pt}}
\put(220,225){\usebox{\plotpoint}}
\put(220,223.67){\rule{7.227pt}{0.400pt}}
\multiput(220.00,224.17)(15.000,-1.000){2}{\rule{3.613pt}{0.400pt}}
\put(250,222.67){\rule{7.227pt}{0.400pt}}
\multiput(250.00,223.17)(15.000,-1.000){2}{\rule{3.613pt}{0.400pt}}
\put(280,221.67){\rule{7.468pt}{0.400pt}}
\multiput(280.00,222.17)(15.500,-1.000){2}{\rule{3.734pt}{0.400pt}}
\put(311,220.67){\rule{7.227pt}{0.400pt}}
\multiput(311.00,221.17)(15.000,-1.000){2}{\rule{3.613pt}{0.400pt}}
\put(341,219.67){\rule{7.227pt}{0.400pt}}
\multiput(341.00,220.17)(15.000,-1.000){2}{\rule{3.613pt}{0.400pt}}
\put(371,218.67){\rule{7.227pt}{0.400pt}}
\multiput(371.00,219.17)(15.000,-1.000){2}{\rule{3.613pt}{0.400pt}}
\put(401,217.67){\rule{7.227pt}{0.400pt}}
\multiput(401.00,218.17)(15.000,-1.000){2}{\rule{3.613pt}{0.400pt}}
\put(431,216.67){\rule{7.468pt}{0.400pt}}
\multiput(431.00,217.17)(15.500,-1.000){2}{\rule{3.734pt}{0.400pt}}
\put(462,215.67){\rule{7.227pt}{0.400pt}}
\multiput(462.00,216.17)(15.000,-1.000){2}{\rule{3.613pt}{0.400pt}}
\put(492,214.67){\rule{7.227pt}{0.400pt}}
\multiput(492.00,215.17)(15.000,-1.000){2}{\rule{3.613pt}{0.400pt}}
\put(522,213.17){\rule{6.100pt}{0.400pt}}
\multiput(522.00,214.17)(17.339,-2.000){2}{\rule{3.050pt}{0.400pt}}
\put(552,211.67){\rule{7.227pt}{0.400pt}}
\multiput(552.00,212.17)(15.000,-1.000){2}{\rule{3.613pt}{0.400pt}}
\put(582,210.17){\rule{6.300pt}{0.400pt}}
\multiput(582.00,211.17)(17.924,-2.000){2}{\rule{3.150pt}{0.400pt}}
\put(613,208.67){\rule{7.227pt}{0.400pt}}
\multiput(613.00,209.17)(15.000,-1.000){2}{\rule{3.613pt}{0.400pt}}
\put(643,207.17){\rule{6.100pt}{0.400pt}}
\multiput(643.00,208.17)(17.339,-2.000){2}{\rule{3.050pt}{0.400pt}}
\put(673,205.67){\rule{7.227pt}{0.400pt}}
\multiput(673.00,206.17)(15.000,-1.000){2}{\rule{3.613pt}{0.400pt}}
\put(703,204.17){\rule{6.100pt}{0.400pt}}
\multiput(703.00,205.17)(17.339,-2.000){2}{\rule{3.050pt}{0.400pt}}
\put(733,202.17){\rule{6.300pt}{0.400pt}}
\multiput(733.00,203.17)(17.924,-2.000){2}{\rule{3.150pt}{0.400pt}}
\put(764,200.17){\rule{6.100pt}{0.400pt}}
\multiput(764.00,201.17)(17.339,-2.000){2}{\rule{3.050pt}{0.400pt}}
\put(794,198.17){\rule{6.100pt}{0.400pt}}
\multiput(794.00,199.17)(17.339,-2.000){2}{\rule{3.050pt}{0.400pt}}
\put(824,196.17){\rule{6.100pt}{0.400pt}}
\multiput(824.00,197.17)(17.339,-2.000){2}{\rule{3.050pt}{0.400pt}}
\put(854,194.17){\rule{6.100pt}{0.400pt}}
\multiput(854.00,195.17)(17.339,-2.000){2}{\rule{3.050pt}{0.400pt}}
\put(884,192.17){\rule{6.300pt}{0.400pt}}
\multiput(884.00,193.17)(17.924,-2.000){2}{\rule{3.150pt}{0.400pt}}
\put(915,190.17){\rule{6.100pt}{0.400pt}}
\multiput(915.00,191.17)(17.339,-2.000){2}{\rule{3.050pt}{0.400pt}}
\multiput(945.00,188.95)(6.490,-0.447){3}{\rule{4.100pt}{0.108pt}}
\multiput(945.00,189.17)(21.490,-3.000){2}{\rule{2.050pt}{0.400pt}}
\put(975,185.17){\rule{6.100pt}{0.400pt}}
\multiput(975.00,186.17)(17.339,-2.000){2}{\rule{3.050pt}{0.400pt}}
\multiput(1005.00,183.95)(6.490,-0.447){3}{\rule{4.100pt}{0.108pt}}
\multiput(1005.00,184.17)(21.490,-3.000){2}{\rule{2.050pt}{0.400pt}}
\multiput(1035.00,180.95)(6.714,-0.447){3}{\rule{4.233pt}{0.108pt}}
\multiput(1035.00,181.17)(22.214,-3.000){2}{\rule{2.117pt}{0.400pt}}
\put(1066,177.17){\rule{6.100pt}{0.400pt}}
\multiput(1066.00,178.17)(17.339,-2.000){2}{\rule{3.050pt}{0.400pt}}
\multiput(1096.00,175.95)(6.490,-0.447){3}{\rule{4.100pt}{0.108pt}}
\multiput(1096.00,176.17)(21.490,-3.000){2}{\rule{2.050pt}{0.400pt}}
\end{picture}

%% file: fig4.tex
\setlength{\unitlength}{0.240900pt}
\ifx\plotpoint\undefined\newsavebox{\plotpoint}\fi
\sbox{\plotpoint}{\rule[-0.200pt]{0.400pt}{0.400pt}}%
\begin{picture}(1190,900)(0,0)
\font\gnuplot=cmr10 at 10pt
\gnuplot
\sbox{\plotpoint}{\rule[-0.200pt]{0.400pt}{0.400pt}}%
\put(220.0,113.0){\rule[-0.200pt]{218.255pt}{0.400pt}}
\put(220.0,113.0){\rule[-0.200pt]{0.400pt}{184.048pt}}
\put(220.0,113.0){\rule[-0.200pt]{4.818pt}{0.400pt}}
\put(198,113){\makebox(0,0)[r]{0}}
\put(1106.0,113.0){\rule[-0.200pt]{4.818pt}{0.400pt}}
\put(220.0,189.0){\rule[-0.200pt]{4.818pt}{0.400pt}}
\put(198,189){\makebox(0,0)[r]{0.02}}
\put(1106.0,189.0){\rule[-0.200pt]{4.818pt}{0.400pt}}
\put(220.0,266.0){\rule[-0.200pt]{4.818pt}{0.400pt}}
\put(198,266){\makebox(0,0)[r]{0.04}}
\put(1106.0,266.0){\rule[-0.200pt]{4.818pt}{0.400pt}}
\put(220.0,342.0){\rule[-0.200pt]{4.818pt}{0.400pt}}
\put(198,342){\makebox(0,0)[r]{0.06}}
\put(1106.0,342.0){\rule[-0.200pt]{4.818pt}{0.400pt}}
\put(220.0,419.0){\rule[-0.200pt]{4.818pt}{0.400pt}}
\put(198,419){\makebox(0,0)[r]{0.08}}
\put(1106.0,419.0){\rule[-0.200pt]{4.818pt}{0.400pt}}
\put(220.0,495.0){\rule[-0.200pt]{4.818pt}{0.400pt}}
\put(198,495){\makebox(0,0)[r]{0.1}}
\put(1106.0,495.0){\rule[-0.200pt]{4.818pt}{0.400pt}}
\put(220.0,571.0){\rule[-0.200pt]{4.818pt}{0.400pt}}
\put(198,571){\makebox(0,0)[r]{0.12}}
\put(1106.0,571.0){\rule[-0.200pt]{4.818pt}{0.400pt}}
\put(220.0,648.0){\rule[-0.200pt]{4.818pt}{0.400pt}}
\put(198,648){\makebox(0,0)[r]{0.14}}
\put(1106.0,648.0){\rule[-0.200pt]{4.818pt}{0.400pt}}
\put(220.0,724.0){\rule[-0.200pt]{4.818pt}{0.400pt}}
\put(198,724){\makebox(0,0)[r]{0.16}}
\put(1106.0,724.0){\rule[-0.200pt]{4.818pt}{0.400pt}}
\put(220.0,801.0){\rule[-0.200pt]{4.818pt}{0.400pt}}
\put(198,801){\makebox(0,0)[r]{0.18}}
\put(1106.0,801.0){\rule[-0.200pt]{4.818pt}{0.400pt}}
\put(220.0,877.0){\rule[-0.200pt]{4.818pt}{0.400pt}}
\put(198,877){\makebox(0,0)[r]{0.2}}
\put(1106.0,877.0){\rule[-0.200pt]{4.818pt}{0.400pt}}
\put(220.0,113.0){\rule[-0.200pt]{0.400pt}{4.818pt}}
\put(220,68){\makebox(0,0){0}}
\put(220.0,857.0){\rule[-0.200pt]{0.400pt}{4.818pt}}
\put(311.0,113.0){\rule[-0.200pt]{0.400pt}{4.818pt}}
\put(311,68){\makebox(0,0){0.1}}
\put(311.0,857.0){\rule[-0.200pt]{0.400pt}{4.818pt}}
\put(401.0,113.0){\rule[-0.200pt]{0.400pt}{4.818pt}}
\put(401,68){\makebox(0,0){0.2}}
\put(401.0,857.0){\rule[-0.200pt]{0.400pt}{4.818pt}}
\put(492.0,113.0){\rule[-0.200pt]{0.400pt}{4.818pt}}
\put(492,68){\makebox(0,0){0.3}}
\put(492.0,857.0){\rule[-0.200pt]{0.400pt}{4.818pt}}
\put(582.0,113.0){\rule[-0.200pt]{0.400pt}{4.818pt}}
\put(582,68){\makebox(0,0){0.4}}
\put(582.0,857.0){\rule[-0.200pt]{0.400pt}{4.818pt}}
\put(673.0,113.0){\rule[-0.200pt]{0.400pt}{4.818pt}}
\put(673,68){\makebox(0,0){0.5}}
\put(673.0,857.0){\rule[-0.200pt]{0.400pt}{4.818pt}}
\put(764.0,113.0){\rule[-0.200pt]{0.400pt}{4.818pt}}
\put(764,68){\makebox(0,0){0.6}}
\put(764.0,857.0){\rule[-0.200pt]{0.400pt}{4.818pt}}
\put(854.0,113.0){\rule[-0.200pt]{0.400pt}{4.818pt}}
\put(854,68){\makebox(0,0){0.7}}
\put(854.0,857.0){\rule[-0.200pt]{0.400pt}{4.818pt}}
\put(945.0,113.0){\rule[-0.200pt]{0.400pt}{4.818pt}}
\put(945,68){\makebox(0,0){0.8}}
\put(945.0,857.0){\rule[-0.200pt]{0.400pt}{4.818pt}}
\put(1035.0,113.0){\rule[-0.200pt]{0.400pt}{4.818pt}}
\put(1035,68){\makebox(0,0){0.9}}
\put(1035.0,857.0){\rule[-0.200pt]{0.400pt}{4.818pt}}
\put(1126.0,113.0){\rule[-0.200pt]{0.400pt}{4.818pt}}
\put(1126,68){\makebox(0,0){1}}
\put(1126.0,857.0){\rule[-0.200pt]{0.400pt}{4.818pt}}
\put(220.0,113.0){\rule[-0.200pt]{218.255pt}{0.400pt}}
\put(1126.0,113.0){\rule[-0.200pt]{0.400pt}{184.048pt}}
\put(220.0,877.0){\rule[-0.200pt]{218.255pt}{0.400pt}}
\put(45,495){\makebox(0,0){$U$}}
\put(673,23){\makebox(0,0){$\epsilon$}}
\put(954,762){\makebox(0,0)[l]{$\enu=3.$}}
\put(791,610){\makebox(0,0)[l]{$\enu=4.$}}
\put(700,495){\makebox(0,0)[l]{$\enu=5.$}}
\put(220.0,113.0){\rule[-0.200pt]{0.400pt}{184.048pt}}
\put(229,134){\usebox{\plotpoint}}
\multiput(229.58,134.00)(0.495,1.122){33}{\rule{0.119pt}{0.989pt}}
\multiput(228.17,134.00)(18.000,37.948){2}{\rule{0.400pt}{0.494pt}}
\multiput(247.58,174.00)(0.495,1.219){31}{\rule{0.119pt}{1.065pt}}
\multiput(246.17,174.00)(17.000,38.790){2}{\rule{0.400pt}{0.532pt}}
\multiput(264.58,215.00)(0.495,1.122){33}{\rule{0.119pt}{0.989pt}}
\multiput(263.17,215.00)(18.000,37.948){2}{\rule{0.400pt}{0.494pt}}
\multiput(282.58,255.00)(0.495,1.159){31}{\rule{0.119pt}{1.018pt}}
\multiput(281.17,255.00)(17.000,36.888){2}{\rule{0.400pt}{0.509pt}}
\multiput(299.58,294.00)(0.495,1.093){33}{\rule{0.119pt}{0.967pt}}
\multiput(298.17,294.00)(18.000,36.994){2}{\rule{0.400pt}{0.483pt}}
\multiput(317.58,333.00)(0.495,1.093){33}{\rule{0.119pt}{0.967pt}}
\multiput(316.17,333.00)(18.000,36.994){2}{\rule{0.400pt}{0.483pt}}
\multiput(335.58,372.00)(0.495,1.099){31}{\rule{0.119pt}{0.971pt}}
\multiput(334.17,372.00)(17.000,34.985){2}{\rule{0.400pt}{0.485pt}}
\multiput(352.58,409.00)(0.495,1.037){33}{\rule{0.119pt}{0.922pt}}
\multiput(351.17,409.00)(18.000,35.086){2}{\rule{0.400pt}{0.461pt}}
\multiput(370.58,446.00)(0.495,1.039){31}{\rule{0.119pt}{0.924pt}}
\multiput(369.17,446.00)(17.000,33.083){2}{\rule{0.400pt}{0.462pt}}
\multiput(387.58,481.00)(0.495,0.952){33}{\rule{0.119pt}{0.856pt}}
\multiput(386.17,481.00)(18.000,32.224){2}{\rule{0.400pt}{0.428pt}}
\multiput(405.58,515.00)(0.495,0.979){31}{\rule{0.119pt}{0.876pt}}
\multiput(404.17,515.00)(17.000,31.181){2}{\rule{0.400pt}{0.438pt}}
\multiput(422.58,548.00)(0.495,0.895){33}{\rule{0.119pt}{0.811pt}}
\multiput(421.17,548.00)(18.000,30.316){2}{\rule{0.400pt}{0.406pt}}
\multiput(440.58,580.00)(0.495,0.838){33}{\rule{0.119pt}{0.767pt}}
\multiput(439.17,580.00)(18.000,28.409){2}{\rule{0.400pt}{0.383pt}}
\multiput(458.58,610.00)(0.495,0.828){31}{\rule{0.119pt}{0.759pt}}
\multiput(457.17,610.00)(17.000,26.425){2}{\rule{0.400pt}{0.379pt}}
\multiput(475.58,638.00)(0.495,0.753){33}{\rule{0.119pt}{0.700pt}}
\multiput(474.17,638.00)(18.000,25.547){2}{\rule{0.400pt}{0.350pt}}
\multiput(493.58,665.00)(0.495,0.738){31}{\rule{0.119pt}{0.688pt}}
\multiput(492.17,665.00)(17.000,23.572){2}{\rule{0.400pt}{0.344pt}}
\multiput(510.58,690.00)(0.495,0.668){33}{\rule{0.119pt}{0.633pt}}
\multiput(509.17,690.00)(18.000,22.685){2}{\rule{0.400pt}{0.317pt}}
\multiput(528.58,714.00)(0.495,0.618){31}{\rule{0.119pt}{0.594pt}}
\multiput(527.17,714.00)(17.000,19.767){2}{\rule{0.400pt}{0.297pt}}
\multiput(545.58,735.00)(0.495,0.554){33}{\rule{0.119pt}{0.544pt}}
\multiput(544.17,735.00)(18.000,18.870){2}{\rule{0.400pt}{0.272pt}}
\multiput(563.00,755.58)(0.498,0.495){33}{\rule{0.500pt}{0.119pt}}
\multiput(563.00,754.17)(16.962,18.000){2}{\rule{0.250pt}{0.400pt}}
\multiput(581.00,773.58)(0.529,0.494){29}{\rule{0.525pt}{0.119pt}}
\multiput(581.00,772.17)(15.910,16.000){2}{\rule{0.263pt}{0.400pt}}
\multiput(598.00,789.58)(0.600,0.494){27}{\rule{0.580pt}{0.119pt}}
\multiput(598.00,788.17)(16.796,15.000){2}{\rule{0.290pt}{0.400pt}}
\multiput(616.00,804.58)(0.712,0.492){21}{\rule{0.667pt}{0.119pt}}
\multiput(616.00,803.17)(15.616,12.000){2}{\rule{0.333pt}{0.400pt}}
\multiput(633.00,816.58)(0.826,0.492){19}{\rule{0.755pt}{0.118pt}}
\multiput(633.00,815.17)(16.434,11.000){2}{\rule{0.377pt}{0.400pt}}
\multiput(651.00,827.59)(1.019,0.489){15}{\rule{0.900pt}{0.118pt}}
\multiput(651.00,826.17)(16.132,9.000){2}{\rule{0.450pt}{0.400pt}}
\multiput(669.00,836.59)(1.255,0.485){11}{\rule{1.071pt}{0.117pt}}
\multiput(669.00,835.17)(14.776,7.000){2}{\rule{0.536pt}{0.400pt}}
\multiput(686.00,843.59)(1.935,0.477){7}{\rule{1.540pt}{0.115pt}}
\multiput(686.00,842.17)(14.804,5.000){2}{\rule{0.770pt}{0.400pt}}
\multiput(704.00,848.60)(2.382,0.468){5}{\rule{1.800pt}{0.113pt}}
\multiput(704.00,847.17)(13.264,4.000){2}{\rule{0.900pt}{0.400pt}}
\put(721,852.17){\rule{3.700pt}{0.400pt}}
\multiput(721.00,851.17)(10.320,2.000){2}{\rule{1.850pt}{0.400pt}}
\put(756,852.17){\rule{3.700pt}{0.400pt}}
\multiput(756.00,853.17)(10.320,-2.000){2}{\rule{1.850pt}{0.400pt}}
\multiput(774.00,850.95)(3.811,-0.447){3}{\rule{2.500pt}{0.108pt}}
\multiput(774.00,851.17)(12.811,-3.000){2}{\rule{1.250pt}{0.400pt}}
\multiput(792.00,847.93)(1.823,-0.477){7}{\rule{1.460pt}{0.115pt}}
\multiput(792.00,848.17)(13.970,-5.000){2}{\rule{0.730pt}{0.400pt}}
\multiput(809.00,842.93)(1.575,-0.482){9}{\rule{1.300pt}{0.116pt}}
\multiput(809.00,843.17)(15.302,-6.000){2}{\rule{0.650pt}{0.400pt}}
\multiput(827.00,836.93)(1.088,-0.488){13}{\rule{0.950pt}{0.117pt}}
\multiput(827.00,837.17)(15.028,-8.000){2}{\rule{0.475pt}{0.400pt}}
\multiput(844.00,828.92)(0.912,-0.491){17}{\rule{0.820pt}{0.118pt}}
\multiput(844.00,829.17)(16.298,-10.000){2}{\rule{0.410pt}{0.400pt}}
\multiput(862.00,818.92)(0.826,-0.492){19}{\rule{0.755pt}{0.118pt}}
\multiput(862.00,819.17)(16.434,-11.000){2}{\rule{0.377pt}{0.400pt}}
\multiput(880.00,807.92)(0.655,-0.493){23}{\rule{0.623pt}{0.119pt}}
\multiput(880.00,808.17)(15.707,-13.000){2}{\rule{0.312pt}{0.400pt}}
\multiput(897.00,794.92)(0.644,-0.494){25}{\rule{0.614pt}{0.119pt}}
\multiput(897.00,795.17)(16.725,-14.000){2}{\rule{0.307pt}{0.400pt}}
\multiput(915.00,780.92)(0.566,-0.494){27}{\rule{0.553pt}{0.119pt}}
\multiput(915.00,781.17)(15.852,-15.000){2}{\rule{0.277pt}{0.400pt}}
\multiput(932.00,765.92)(0.498,-0.495){33}{\rule{0.500pt}{0.119pt}}
\multiput(932.00,766.17)(16.962,-18.000){2}{\rule{0.250pt}{0.400pt}}
\multiput(950.58,746.73)(0.495,-0.558){31}{\rule{0.119pt}{0.547pt}}
\multiput(949.17,747.86)(17.000,-17.865){2}{\rule{0.400pt}{0.274pt}}
\multiput(967.58,727.74)(0.495,-0.554){33}{\rule{0.119pt}{0.544pt}}
\multiput(966.17,728.87)(18.000,-18.870){2}{\rule{0.400pt}{0.272pt}}
\multiput(985.58,707.46)(0.495,-0.639){33}{\rule{0.119pt}{0.611pt}}
\multiput(984.17,708.73)(18.000,-21.732){2}{\rule{0.400pt}{0.306pt}}
\multiput(1003.58,684.14)(0.495,-0.738){31}{\rule{0.119pt}{0.688pt}}
\multiput(1002.17,685.57)(17.000,-23.572){2}{\rule{0.400pt}{0.344pt}}
\multiput(1020.58,659.09)(0.495,-0.753){33}{\rule{0.119pt}{0.700pt}}
\multiput(1019.17,660.55)(18.000,-25.547){2}{\rule{0.400pt}{0.350pt}}
\multiput(1038.58,631.65)(0.495,-0.888){31}{\rule{0.119pt}{0.806pt}}
\multiput(1037.17,633.33)(17.000,-28.327){2}{\rule{0.400pt}{0.403pt}}
\multiput(1055.58,601.36)(0.495,-0.980){33}{\rule{0.119pt}{0.878pt}}
\multiput(1054.17,603.18)(18.000,-33.178){2}{\rule{0.400pt}{0.439pt}}
\multiput(1073.58,565.58)(0.495,-1.219){31}{\rule{0.119pt}{1.065pt}}
\multiput(1072.17,567.79)(17.000,-38.790){2}{\rule{0.400pt}{0.532pt}}
\multiput(1090.58,523.60)(0.495,-1.519){33}{\rule{0.119pt}{1.300pt}}
\multiput(1089.17,526.30)(18.000,-51.302){2}{\rule{0.400pt}{0.650pt}}
\put(739.0,854.0){\rule[-0.200pt]{4.095pt}{0.400pt}}
\put(229,134){\usebox{\plotpoint}}
\multiput(229.58,134.00)(0.495,1.122){33}{\rule{0.119pt}{0.989pt}}
\multiput(228.17,134.00)(18.000,37.948){2}{\rule{0.400pt}{0.494pt}}
\multiput(247.58,174.00)(0.495,1.219){31}{\rule{0.119pt}{1.065pt}}
\multiput(246.17,174.00)(17.000,38.790){2}{\rule{0.400pt}{0.532pt}}
\multiput(264.58,215.00)(0.495,1.122){33}{\rule{0.119pt}{0.989pt}}
\multiput(263.17,215.00)(18.000,37.948){2}{\rule{0.400pt}{0.494pt}}
\multiput(282.58,255.00)(0.495,1.159){31}{\rule{0.119pt}{1.018pt}}
\multiput(281.17,255.00)(17.000,36.888){2}{\rule{0.400pt}{0.509pt}}
\multiput(299.58,294.00)(0.495,1.065){33}{\rule{0.119pt}{0.944pt}}
\multiput(298.17,294.00)(18.000,36.040){2}{\rule{0.400pt}{0.472pt}}
\multiput(317.58,332.00)(0.495,1.037){33}{\rule{0.119pt}{0.922pt}}
\multiput(316.17,332.00)(18.000,35.086){2}{\rule{0.400pt}{0.461pt}}
\multiput(335.58,369.00)(0.495,1.039){31}{\rule{0.119pt}{0.924pt}}
\multiput(334.17,369.00)(17.000,33.083){2}{\rule{0.400pt}{0.462pt}}
\multiput(352.58,404.00)(0.495,0.952){33}{\rule{0.119pt}{0.856pt}}
\multiput(351.17,404.00)(18.000,32.224){2}{\rule{0.400pt}{0.428pt}}
\multiput(370.58,438.00)(0.495,0.949){31}{\rule{0.119pt}{0.853pt}}
\multiput(369.17,438.00)(17.000,30.230){2}{\rule{0.400pt}{0.426pt}}
\multiput(387.58,470.00)(0.495,0.866){33}{\rule{0.119pt}{0.789pt}}
\multiput(386.17,470.00)(18.000,29.363){2}{\rule{0.400pt}{0.394pt}}
\multiput(405.58,501.00)(0.495,0.828){31}{\rule{0.119pt}{0.759pt}}
\multiput(404.17,501.00)(17.000,26.425){2}{\rule{0.400pt}{0.379pt}}
\multiput(422.58,529.00)(0.495,0.725){33}{\rule{0.119pt}{0.678pt}}
\multiput(421.17,529.00)(18.000,24.593){2}{\rule{0.400pt}{0.339pt}}
\multiput(440.58,555.00)(0.495,0.668){33}{\rule{0.119pt}{0.633pt}}
\multiput(439.17,555.00)(18.000,22.685){2}{\rule{0.400pt}{0.317pt}}
\multiput(458.58,579.00)(0.495,0.648){31}{\rule{0.119pt}{0.618pt}}
\multiput(457.17,579.00)(17.000,20.718){2}{\rule{0.400pt}{0.309pt}}
\multiput(475.58,601.00)(0.495,0.526){33}{\rule{0.119pt}{0.522pt}}
\multiput(474.17,601.00)(18.000,17.916){2}{\rule{0.400pt}{0.261pt}}
\multiput(493.00,620.58)(0.497,0.495){31}{\rule{0.500pt}{0.119pt}}
\multiput(493.00,619.17)(15.962,17.000){2}{\rule{0.250pt}{0.400pt}}
\multiput(510.00,637.58)(0.600,0.494){27}{\rule{0.580pt}{0.119pt}}
\multiput(510.00,636.17)(16.796,15.000){2}{\rule{0.290pt}{0.400pt}}
\multiput(528.00,652.58)(0.712,0.492){21}{\rule{0.667pt}{0.119pt}}
\multiput(528.00,651.17)(15.616,12.000){2}{\rule{0.333pt}{0.400pt}}
\multiput(545.00,664.58)(0.912,0.491){17}{\rule{0.820pt}{0.118pt}}
\multiput(545.00,663.17)(16.298,10.000){2}{\rule{0.410pt}{0.400pt}}
\multiput(563.00,674.59)(1.332,0.485){11}{\rule{1.129pt}{0.117pt}}
\multiput(563.00,673.17)(15.658,7.000){2}{\rule{0.564pt}{0.400pt}}
\multiput(581.00,681.59)(1.823,0.477){7}{\rule{1.460pt}{0.115pt}}
\multiput(581.00,680.17)(13.970,5.000){2}{\rule{0.730pt}{0.400pt}}
\multiput(598.00,686.61)(3.811,0.447){3}{\rule{2.500pt}{0.108pt}}
\multiput(598.00,685.17)(12.811,3.000){2}{\rule{1.250pt}{0.400pt}}
\put(633,687.17){\rule{3.700pt}{0.400pt}}
\multiput(633.00,688.17)(10.320,-2.000){2}{\rule{1.850pt}{0.400pt}}
\multiput(651.00,685.94)(2.528,-0.468){5}{\rule{1.900pt}{0.113pt}}
\multiput(651.00,686.17)(14.056,-4.000){2}{\rule{0.950pt}{0.400pt}}
\multiput(669.00,681.93)(1.255,-0.485){11}{\rule{1.071pt}{0.117pt}}
\multiput(669.00,682.17)(14.776,-7.000){2}{\rule{0.536pt}{0.400pt}}
\multiput(686.00,674.93)(1.019,-0.489){15}{\rule{0.900pt}{0.118pt}}
\multiput(686.00,675.17)(16.132,-9.000){2}{\rule{0.450pt}{0.400pt}}
\multiput(704.00,665.92)(0.779,-0.492){19}{\rule{0.718pt}{0.118pt}}
\multiput(704.00,666.17)(15.509,-11.000){2}{\rule{0.359pt}{0.400pt}}
\multiput(721.00,654.92)(0.644,-0.494){25}{\rule{0.614pt}{0.119pt}}
\multiput(721.00,655.17)(16.725,-14.000){2}{\rule{0.307pt}{0.400pt}}
\multiput(739.00,640.92)(0.529,-0.494){29}{\rule{0.525pt}{0.119pt}}
\multiput(739.00,641.17)(15.910,-16.000){2}{\rule{0.263pt}{0.400pt}}
\multiput(756.00,624.92)(0.498,-0.495){33}{\rule{0.500pt}{0.119pt}}
\multiput(756.00,625.17)(16.962,-18.000){2}{\rule{0.250pt}{0.400pt}}
\multiput(774.58,605.65)(0.495,-0.583){33}{\rule{0.119pt}{0.567pt}}
\multiput(773.17,606.82)(18.000,-19.824){2}{\rule{0.400pt}{0.283pt}}
\multiput(792.58,584.24)(0.495,-0.708){31}{\rule{0.119pt}{0.665pt}}
\multiput(791.17,585.62)(17.000,-22.620){2}{\rule{0.400pt}{0.332pt}}
\multiput(809.58,560.09)(0.495,-0.753){33}{\rule{0.119pt}{0.700pt}}
\multiput(808.17,561.55)(18.000,-25.547){2}{\rule{0.400pt}{0.350pt}}
\multiput(827.58,532.65)(0.495,-0.888){31}{\rule{0.119pt}{0.806pt}}
\multiput(826.17,534.33)(17.000,-28.327){2}{\rule{0.400pt}{0.403pt}}
\multiput(844.58,502.36)(0.495,-0.980){33}{\rule{0.119pt}{0.878pt}}
\multiput(843.17,504.18)(18.000,-33.178){2}{\rule{0.400pt}{0.439pt}}
\multiput(862.58,466.90)(0.495,-1.122){33}{\rule{0.119pt}{0.989pt}}
\multiput(861.17,468.95)(18.000,-37.948){2}{\rule{0.400pt}{0.494pt}}
\multiput(880.58,425.80)(0.495,-1.460){31}{\rule{0.119pt}{1.253pt}}
\multiput(879.17,428.40)(17.000,-46.399){2}{\rule{0.400pt}{0.626pt}}
\multiput(897.58,375.87)(0.495,-1.746){33}{\rule{0.119pt}{1.478pt}}
\multiput(896.17,378.93)(18.000,-58.933){2}{\rule{0.400pt}{0.739pt}}
\multiput(915.58,310.01)(0.495,-2.934){31}{\rule{0.119pt}{2.406pt}}
\multiput(914.17,315.01)(17.000,-93.006){2}{\rule{0.400pt}{1.203pt}}
\multiput(932.58,211.53)(0.495,-3.079){33}{\rule{0.119pt}{2.522pt}}
\multiput(931.17,216.77)(18.000,-103.765){2}{\rule{0.400pt}{1.261pt}}
\put(616.0,689.0){\rule[-0.200pt]{4.095pt}{0.400pt}}
\put(950.0,113.0){\rule[-0.200pt]{38.062pt}{0.400pt}}
\put(229,134){\usebox{\plotpoint}}
\multiput(229.58,134.00)(0.495,1.150){33}{\rule{0.119pt}{1.011pt}}
\multiput(228.17,134.00)(18.000,38.901){2}{\rule{0.400pt}{0.506pt}}
\multiput(247.58,175.00)(0.495,1.189){31}{\rule{0.119pt}{1.041pt}}
\multiput(246.17,175.00)(17.000,37.839){2}{\rule{0.400pt}{0.521pt}}
\multiput(264.58,215.00)(0.495,1.093){33}{\rule{0.119pt}{0.967pt}}
\multiput(263.17,215.00)(18.000,36.994){2}{\rule{0.400pt}{0.483pt}}
\multiput(282.58,254.00)(0.495,1.159){31}{\rule{0.119pt}{1.018pt}}
\multiput(281.17,254.00)(17.000,36.888){2}{\rule{0.400pt}{0.509pt}}
\multiput(299.58,293.00)(0.495,1.037){33}{\rule{0.119pt}{0.922pt}}
\multiput(298.17,293.00)(18.000,35.086){2}{\rule{0.400pt}{0.461pt}}
\multiput(317.58,330.00)(0.495,0.980){33}{\rule{0.119pt}{0.878pt}}
\multiput(316.17,330.00)(18.000,33.178){2}{\rule{0.400pt}{0.439pt}}
\multiput(335.58,365.00)(0.495,0.979){31}{\rule{0.119pt}{0.876pt}}
\multiput(334.17,365.00)(17.000,31.181){2}{\rule{0.400pt}{0.438pt}}
\multiput(352.58,398.00)(0.495,0.866){33}{\rule{0.119pt}{0.789pt}}
\multiput(351.17,398.00)(18.000,29.363){2}{\rule{0.400pt}{0.394pt}}
\multiput(370.58,429.00)(0.495,0.858){31}{\rule{0.119pt}{0.782pt}}
\multiput(369.17,429.00)(17.000,27.376){2}{\rule{0.400pt}{0.391pt}}
\multiput(387.58,458.00)(0.495,0.725){33}{\rule{0.119pt}{0.678pt}}
\multiput(386.17,458.00)(18.000,24.593){2}{\rule{0.400pt}{0.339pt}}
\multiput(405.58,484.00)(0.495,0.678){31}{\rule{0.119pt}{0.641pt}}
\multiput(404.17,484.00)(17.000,21.669){2}{\rule{0.400pt}{0.321pt}}
\multiput(422.58,507.00)(0.495,0.554){33}{\rule{0.119pt}{0.544pt}}
\multiput(421.17,507.00)(18.000,18.870){2}{\rule{0.400pt}{0.272pt}}
\multiput(440.00,527.58)(0.498,0.495){33}{\rule{0.500pt}{0.119pt}}
\multiput(440.00,526.17)(16.962,18.000){2}{\rule{0.250pt}{0.400pt}}
\multiput(458.00,545.58)(0.607,0.494){25}{\rule{0.586pt}{0.119pt}}
\multiput(458.00,544.17)(15.784,14.000){2}{\rule{0.293pt}{0.400pt}}
\multiput(475.00,559.58)(0.826,0.492){19}{\rule{0.755pt}{0.118pt}}
\multiput(475.00,558.17)(16.434,11.000){2}{\rule{0.377pt}{0.400pt}}
\multiput(493.00,570.59)(0.961,0.489){15}{\rule{0.856pt}{0.118pt}}
\multiput(493.00,569.17)(15.224,9.000){2}{\rule{0.428pt}{0.400pt}}
\multiput(510.00,579.59)(1.935,0.477){7}{\rule{1.540pt}{0.115pt}}
\multiput(510.00,578.17)(14.804,5.000){2}{\rule{0.770pt}{0.400pt}}
\put(528,584.17){\rule{3.500pt}{0.400pt}}
\multiput(528.00,583.17)(9.736,2.000){2}{\rule{1.750pt}{0.400pt}}
\put(545,584.67){\rule{4.336pt}{0.400pt}}
\multiput(545.00,585.17)(9.000,-1.000){2}{\rule{2.168pt}{0.400pt}}
\multiput(563.00,583.93)(1.935,-0.477){7}{\rule{1.540pt}{0.115pt}}
\multiput(563.00,584.17)(14.804,-5.000){2}{\rule{0.770pt}{0.400pt}}
\multiput(581.00,578.93)(1.255,-0.485){11}{\rule{1.071pt}{0.117pt}}
\multiput(581.00,579.17)(14.776,-7.000){2}{\rule{0.536pt}{0.400pt}}
\multiput(598.00,571.92)(0.826,-0.492){19}{\rule{0.755pt}{0.118pt}}
\multiput(598.00,572.17)(16.434,-11.000){2}{\rule{0.377pt}{0.400pt}}
\multiput(616.00,560.92)(0.607,-0.494){25}{\rule{0.586pt}{0.119pt}}
\multiput(616.00,561.17)(15.784,-14.000){2}{\rule{0.293pt}{0.400pt}}
\multiput(633.00,546.92)(0.528,-0.495){31}{\rule{0.524pt}{0.119pt}}
\multiput(633.00,547.17)(16.913,-17.000){2}{\rule{0.262pt}{0.400pt}}
\multiput(651.58,528.65)(0.495,-0.583){33}{\rule{0.119pt}{0.567pt}}
\multiput(650.17,529.82)(18.000,-19.824){2}{\rule{0.400pt}{0.283pt}}
\multiput(669.58,507.14)(0.495,-0.738){31}{\rule{0.119pt}{0.688pt}}
\multiput(668.17,508.57)(17.000,-23.572){2}{\rule{0.400pt}{0.344pt}}
\multiput(686.58,481.82)(0.495,-0.838){33}{\rule{0.119pt}{0.767pt}}
\multiput(685.17,483.41)(18.000,-28.409){2}{\rule{0.400pt}{0.383pt}}
\multiput(704.58,451.17)(0.495,-1.039){31}{\rule{0.119pt}{0.924pt}}
\multiput(703.17,453.08)(17.000,-33.083){2}{\rule{0.400pt}{0.462pt}}
\multiput(721.58,415.62)(0.495,-1.207){33}{\rule{0.119pt}{1.056pt}}
\multiput(720.17,417.81)(18.000,-40.809){2}{\rule{0.400pt}{0.528pt}}
\multiput(739.58,371.21)(0.495,-1.640){31}{\rule{0.119pt}{1.394pt}}
\multiput(738.17,374.11)(17.000,-52.106){2}{\rule{0.400pt}{0.697pt}}
\multiput(756.58,314.30)(0.495,-2.228){33}{\rule{0.119pt}{1.856pt}}
\multiput(755.17,318.15)(18.000,-75.149){2}{\rule{0.400pt}{0.928pt}}
\multiput(774.58,230.59)(0.495,-3.675){33}{\rule{0.119pt}{2.989pt}}
\multiput(773.17,236.80)(18.000,-123.796){2}{\rule{0.400pt}{1.494pt}}
\put(792.0,113.0){\rule[-0.200pt]{76.124pt}{0.400pt}}
\end{picture}

%% file: fig5.tex
\setlength{\unitlength}{0.240900pt}
\ifx\plotpoint\undefined\newsavebox{\plotpoint}\fi
\sbox{\plotpoint}{\rule[-0.200pt]{0.400pt}{0.400pt}}%
\begin{picture}(1190,900)(0,0)
\font\gnuplot=cmr10 at 10pt
\gnuplot
\sbox{\plotpoint}{\rule[-0.200pt]{0.400pt}{0.400pt}}%
\put(220.0,113.0){\rule[-0.200pt]{218.255pt}{0.400pt}}
\put(220.0,113.0){\rule[-0.200pt]{4.818pt}{0.400pt}}
\put(198,113){\makebox(0,0)[r]{0}}
\put(1106.0,113.0){\rule[-0.200pt]{4.818pt}{0.400pt}}
\put(220.0,189.0){\rule[-0.200pt]{4.818pt}{0.400pt}}
\put(198,189){\makebox(0,0)[r]{0.05}}
\put(1106.0,189.0){\rule[-0.200pt]{4.818pt}{0.400pt}}
\put(220.0,266.0){\rule[-0.200pt]{4.818pt}{0.400pt}}
\put(198,266){\makebox(0,0)[r]{0.1}}
\put(1106.0,266.0){\rule[-0.200pt]{4.818pt}{0.400pt}}
\put(220.0,342.0){\rule[-0.200pt]{4.818pt}{0.400pt}}
\put(198,342){\makebox(0,0)[r]{0.15}}
\put(1106.0,342.0){\rule[-0.200pt]{4.818pt}{0.400pt}}
\put(220.0,419.0){\rule[-0.200pt]{4.818pt}{0.400pt}}
\put(198,419){\makebox(0,0)[r]{0.2}}
\put(1106.0,419.0){\rule[-0.200pt]{4.818pt}{0.400pt}}
\put(220.0,495.0){\rule[-0.200pt]{4.818pt}{0.400pt}}
\put(198,495){\makebox(0,0)[r]{0.25}}
\put(1106.0,495.0){\rule[-0.200pt]{4.818pt}{0.400pt}}
\put(220.0,571.0){\rule[-0.200pt]{4.818pt}{0.400pt}}
\put(198,571){\makebox(0,0)[r]{0.3}}
\put(1106.0,571.0){\rule[-0.200pt]{4.818pt}{0.400pt}}
\put(220.0,648.0){\rule[-0.200pt]{4.818pt}{0.400pt}}
\put(198,648){\makebox(0,0)[r]{0.35}}
\put(1106.0,648.0){\rule[-0.200pt]{4.818pt}{0.400pt}}
\put(220.0,724.0){\rule[-0.200pt]{4.818pt}{0.400pt}}
\put(198,724){\makebox(0,0)[r]{0.4}}
\put(1106.0,724.0){\rule[-0.200pt]{4.818pt}{0.400pt}}
\put(220.0,801.0){\rule[-0.200pt]{4.818pt}{0.400pt}}
\put(198,801){\makebox(0,0)[r]{0.45}}
\put(1106.0,801.0){\rule[-0.200pt]{4.818pt}{0.400pt}}
\put(220.0,877.0){\rule[-0.200pt]{4.818pt}{0.400pt}}
\put(198,877){\makebox(0,0)[r]{0.5}}
\put(1106.0,877.0){\rule[-0.200pt]{4.818pt}{0.400pt}}
\put(220.0,113.0){\rule[-0.200pt]{0.400pt}{4.818pt}}
\put(220,68){\makebox(0,0){0.2}}
\put(220.0,857.0){\rule[-0.200pt]{0.400pt}{4.818pt}}
\put(333.0,113.0){\rule[-0.200pt]{0.400pt}{4.818pt}}
\put(333,68){\makebox(0,0){0.3}}
\put(333.0,857.0){\rule[-0.200pt]{0.400pt}{4.818pt}}
\put(447.0,113.0){\rule[-0.200pt]{0.400pt}{4.818pt}}
\put(447,68){\makebox(0,0){0.4}}
\put(447.0,857.0){\rule[-0.200pt]{0.400pt}{4.818pt}}
\put(560.0,113.0){\rule[-0.200pt]{0.400pt}{4.818pt}}
\put(560,68){\makebox(0,0){0.5}}
\put(560.0,857.0){\rule[-0.200pt]{0.400pt}{4.818pt}}
\put(673.0,113.0){\rule[-0.200pt]{0.400pt}{4.818pt}}
\put(673,68){\makebox(0,0){0.6}}
\put(673.0,857.0){\rule[-0.200pt]{0.400pt}{4.818pt}}
\put(786.0,113.0){\rule[-0.200pt]{0.400pt}{4.818pt}}
\put(786,68){\makebox(0,0){0.7}}
\put(786.0,857.0){\rule[-0.200pt]{0.400pt}{4.818pt}}
\put(899.0,113.0){\rule[-0.200pt]{0.400pt}{4.818pt}}
\put(899,68){\makebox(0,0){0.8}}
\put(899.0,857.0){\rule[-0.200pt]{0.400pt}{4.818pt}}
\put(1013.0,113.0){\rule[-0.200pt]{0.400pt}{4.818pt}}
\put(1013,68){\makebox(0,0){0.9}}
\put(1013.0,857.0){\rule[-0.200pt]{0.400pt}{4.818pt}}
\put(1126.0,113.0){\rule[-0.200pt]{0.400pt}{4.818pt}}
\put(1126,68){\makebox(0,0){1}}
\put(1126.0,857.0){\rule[-0.200pt]{0.400pt}{4.818pt}}
\put(220.0,113.0){\rule[-0.200pt]{218.255pt}{0.400pt}}
\put(1126.0,113.0){\rule[-0.200pt]{0.400pt}{184.048pt}}
\put(220.0,877.0){\rule[-0.200pt]{218.255pt}{0.400pt}}
\put(45,495){\makebox(0,0){$U$}}
\put(673,23){\makebox(0,0){$r_2/r_1$}}
\put(730,755){\makebox(0,0)[l]{$\enu=1.$}}
\put(730,495){\makebox(0,0)[l]{$\enu=2.$}}
\put(786,251){\makebox(0,0)[l]{$\enu=3.$}}
\put(220.0,113.0){\rule[-0.200pt]{0.400pt}{184.048pt}}
\put(317,585){\usebox{\plotpoint}}
\multiput(317.00,585.61)(3.141,0.447){3}{\rule{2.100pt}{0.108pt}}
\multiput(317.00,584.17)(10.641,3.000){2}{\rule{1.050pt}{0.400pt}}
\multiput(332.00,588.61)(3.365,0.447){3}{\rule{2.233pt}{0.108pt}}
\multiput(332.00,587.17)(11.365,3.000){2}{\rule{1.117pt}{0.400pt}}
\multiput(348.00,591.61)(3.365,0.447){3}{\rule{2.233pt}{0.108pt}}
\multiput(348.00,590.17)(11.365,3.000){2}{\rule{1.117pt}{0.400pt}}
\multiput(364.00,594.61)(3.588,0.447){3}{\rule{2.367pt}{0.108pt}}
\multiput(364.00,593.17)(12.088,3.000){2}{\rule{1.183pt}{0.400pt}}
\multiput(381.00,597.60)(2.674,0.468){5}{\rule{2.000pt}{0.113pt}}
\multiput(381.00,596.17)(14.849,4.000){2}{\rule{1.000pt}{0.400pt}}
\multiput(400.00,601.59)(2.157,0.477){7}{\rule{1.700pt}{0.115pt}}
\multiput(400.00,600.17)(16.472,5.000){2}{\rule{0.850pt}{0.400pt}}
\multiput(420.00,606.60)(2.674,0.468){5}{\rule{2.000pt}{0.113pt}}
\multiput(420.00,605.17)(14.849,4.000){2}{\rule{1.000pt}{0.400pt}}
\multiput(439.00,610.59)(1.937,0.482){9}{\rule{1.567pt}{0.116pt}}
\multiput(439.00,609.17)(18.748,6.000){2}{\rule{0.783pt}{0.400pt}}
\multiput(461.00,616.59)(2.027,0.482){9}{\rule{1.633pt}{0.116pt}}
\multiput(461.00,615.17)(19.610,6.000){2}{\rule{0.817pt}{0.400pt}}
\multiput(484.00,622.59)(1.748,0.488){13}{\rule{1.450pt}{0.117pt}}
\multiput(484.00,621.17)(23.990,8.000){2}{\rule{0.725pt}{0.400pt}}
\multiput(511.00,630.59)(1.865,0.485){11}{\rule{1.529pt}{0.117pt}}
\multiput(511.00,629.17)(21.827,7.000){2}{\rule{0.764pt}{0.400pt}}
\multiput(536.00,637.59)(1.776,0.489){15}{\rule{1.478pt}{0.118pt}}
\multiput(536.00,636.17)(27.933,9.000){2}{\rule{0.739pt}{0.400pt}}
\multiput(567.00,646.58)(1.534,0.492){19}{\rule{1.300pt}{0.118pt}}
\multiput(567.00,645.17)(30.302,11.000){2}{\rule{0.650pt}{0.400pt}}
\multiput(600.00,657.58)(1.581,0.492){19}{\rule{1.336pt}{0.118pt}}
\multiput(600.00,656.17)(31.226,11.000){2}{\rule{0.668pt}{0.400pt}}
\multiput(634.00,668.58)(1.765,0.493){23}{\rule{1.485pt}{0.119pt}}
\multiput(634.00,667.17)(41.919,13.000){2}{\rule{0.742pt}{0.400pt}}
\multiput(679.00,681.58)(1.554,0.494){29}{\rule{1.325pt}{0.119pt}}
\multiput(679.00,680.17)(46.250,16.000){2}{\rule{0.663pt}{0.400pt}}
\multiput(728.00,697.58)(1.568,0.496){37}{\rule{1.340pt}{0.119pt}}
\multiput(728.00,696.17)(59.219,20.000){2}{\rule{0.670pt}{0.400pt}}
\multiput(790.00,717.58)(1.688,0.497){49}{\rule{1.438pt}{0.120pt}}
\multiput(790.00,716.17)(84.014,26.000){2}{\rule{0.719pt}{0.400pt}}
\multiput(877.00,743.58)(1.865,0.499){131}{\rule{1.587pt}{0.120pt}}
\multiput(877.00,742.17)(245.707,67.000){2}{\rule{0.793pt}{0.400pt}}
\put(317,364){\usebox{\plotpoint}}
\multiput(317.00,362.95)(3.141,-0.447){3}{\rule{2.100pt}{0.108pt}}
\multiput(317.00,363.17)(10.641,-3.000){2}{\rule{1.050pt}{0.400pt}}
\put(332,359.67){\rule{3.854pt}{0.400pt}}
\multiput(332.00,360.17)(8.000,-1.000){2}{\rule{1.927pt}{0.400pt}}
\put(348,358.67){\rule{3.854pt}{0.400pt}}
\multiput(348.00,359.17)(8.000,-1.000){2}{\rule{1.927pt}{0.400pt}}
\put(364,357.67){\rule{4.095pt}{0.400pt}}
\multiput(364.00,358.17)(8.500,-1.000){2}{\rule{2.048pt}{0.400pt}}
\put(381,358.17){\rule{3.900pt}{0.400pt}}
\multiput(381.00,357.17)(10.905,2.000){2}{\rule{1.950pt}{0.400pt}}
\put(400,359.67){\rule{4.818pt}{0.400pt}}
\multiput(400.00,359.17)(10.000,1.000){2}{\rule{2.409pt}{0.400pt}}
\put(420,361.17){\rule{3.900pt}{0.400pt}}
\multiput(420.00,360.17)(10.905,2.000){2}{\rule{1.950pt}{0.400pt}}
\multiput(439.00,363.61)(4.704,0.447){3}{\rule{3.033pt}{0.108pt}}
\multiput(439.00,362.17)(15.704,3.000){2}{\rule{1.517pt}{0.400pt}}
\multiput(461.00,366.60)(3.259,0.468){5}{\rule{2.400pt}{0.113pt}}
\multiput(461.00,365.17)(18.019,4.000){2}{\rule{1.200pt}{0.400pt}}
\multiput(484.00,370.59)(2.936,0.477){7}{\rule{2.260pt}{0.115pt}}
\multiput(484.00,369.17)(22.309,5.000){2}{\rule{1.130pt}{0.400pt}}
\multiput(511.00,375.60)(3.552,0.468){5}{\rule{2.600pt}{0.113pt}}
\multiput(511.00,374.17)(19.604,4.000){2}{\rule{1.300pt}{0.400pt}}
\multiput(536.00,379.59)(2.013,0.488){13}{\rule{1.650pt}{0.117pt}}
\multiput(536.00,378.17)(27.575,8.000){2}{\rule{0.825pt}{0.400pt}}
\multiput(567.00,387.59)(2.476,0.485){11}{\rule{1.986pt}{0.117pt}}
\multiput(567.00,386.17)(28.879,7.000){2}{\rule{0.993pt}{0.400pt}}
\multiput(600.00,394.59)(1.951,0.489){15}{\rule{1.611pt}{0.118pt}}
\multiput(600.00,393.17)(30.656,9.000){2}{\rule{0.806pt}{0.400pt}}
\multiput(634.00,403.58)(1.918,0.492){21}{\rule{1.600pt}{0.119pt}}
\multiput(634.00,402.17)(41.679,12.000){2}{\rule{0.800pt}{0.400pt}}
\multiput(679.00,415.58)(1.782,0.494){25}{\rule{1.500pt}{0.119pt}}
\multiput(679.00,414.17)(45.887,14.000){2}{\rule{0.750pt}{0.400pt}}
\multiput(728.00,429.58)(1.851,0.495){31}{\rule{1.559pt}{0.119pt}}
\multiput(728.00,428.17)(58.765,17.000){2}{\rule{0.779pt}{0.400pt}}
\multiput(790.00,446.58)(1.757,0.497){47}{\rule{1.492pt}{0.120pt}}
\multiput(790.00,445.17)(83.903,25.000){2}{\rule{0.746pt}{0.400pt}}
\multiput(877.00,471.58)(1.784,0.499){137}{\rule{1.523pt}{0.120pt}}
\multiput(877.00,470.17)(245.839,70.000){2}{\rule{0.761pt}{0.400pt}}
\put(317,113){\usebox{\plotpoint}}
\multiput(679.00,113.58)(1.782,0.494){25}{\rule{1.500pt}{0.119pt}}
\multiput(679.00,112.17)(45.887,14.000){2}{\rule{0.750pt}{0.400pt}}
\multiput(728.00,127.58)(0.596,0.498){101}{\rule{0.577pt}{0.120pt}}
\multiput(728.00,126.17)(60.803,52.000){2}{\rule{0.288pt}{0.400pt}}
\multiput(790.00,179.58)(1.015,0.498){83}{\rule{0.909pt}{0.120pt}}
\multiput(790.00,178.17)(85.113,43.000){2}{\rule{0.455pt}{0.400pt}}
\multiput(877.00,222.58)(1.341,0.499){183}{\rule{1.171pt}{0.120pt}}
\multiput(877.00,221.17)(246.570,93.000){2}{\rule{0.585pt}{0.400pt}}
\put(317.0,113.0){\rule[-0.200pt]{87.206pt}{0.400pt}}
\end{picture}